\documentstyle[12pt]{article}

\newcommand{\pa}{\partial}
\newcommand{\hf}{\frac12}
\newcommand{\beq}{\begin{equation}}
\newcommand{\eeq}{\end{equation}}
\newcommand{\beqa}{\begin{eqnarray}}
\newcommand{\eeqa}{\end{eqnarray}}
\newcommand{\sigr}{\sigma_{\rm r}}
\newcommand{\sigt}{\sigma_{\rm t}}
\newcommand{\tr}{t_{\rm r}}
\newenvironment{reference}{\bigskip\bigskip\leftline{\Large\bf
References}\nopagebreak \begin{list}{}{\itemindent-\leftmargin \itemsep=0pt
\parsep=0pt}}{\end{list}}
\newcommand{\apj}{ApJ}
\newcommand{\mn}{MNRAS}
\newcommand{\pasj}{PASJ}
\newcommand{\aap}{A\&A}

\title{
\begin{flushright}
\normalsize
OU-TAP-23 \\
submitted to PASJ \\
July 5, 1995
\end{flushright}
\ \\
Fokker-Planck Models of Star Clusters \\
with Anisotropic Velocity Distributions. \\
I. Pre-Collapse Evolution
}
\author{
Koji Takahashi
\thanks{Research Fellow of the Japan Society for the Promotion of Science}\\
{\it Department of Earth and Space Science,} \\
{\it Faculty of Science, Osaka University,}\\
{\it Toyonaka, Osaka 560, Japan} \\
{\it E-mail takahasi@vega.ess.sci.osaka-u.ac.jp}
}
\date{}

\begin{document}

\maketitle

\begin{abstract}

The evolution of a spherical single-mass star cluster is followed in detail up
to core collapse by numerically solving the orbit-averaged two-dimensional
Fokker-Planck equation in energy-angular momentum space.
Velocity anisotropy is allowed in the two-dimensional Fokker-Planck model.
Using improved numerical codes, the evolution has been followed until the
central density increased by a factor of $10^{14}$ with high numerical
accuracy.
The numerical results clearly show self-similar evolution of the core during
the late stages of the core collapse.
In the self-similar region between the isothermal core and the outer halo, the
density profile is characterized by a power law $\rho(r) \propto r^{-2.23}$ and
the ratio of the one-dimensional tangential velocity dispersion to the radial
one is $\sigt^2/\sigr^2=0.92$.
As the core collapse proceeds, the collapse rate $\xi \equiv
\tr(0)d\ln\rho(0)/dt$ tends to the limiting value of $\xi=2.9 \times 10^{-3}$,
which is 19\% smaller than the value for isotropic clusters.
When Plummer's model is chosen as the initial condition,
the collapse time is about 17.6 times the initial half-mass relaxation time.
As the result of strong relaxation in the core, the halo becomes to be
dominated by radial orbits.
The degree of anisotropy monotonically increases as the radius increases.
In the outer halo, the profiles of the density are approximated by $\rho
\propto r^{-3.5}$.
This work confirms that the generation of velocity anisotropy is an important
process in collisional stellar systems.


\end{abstract}

\section{Introduction}\label{sec:intro}

\indent\indent

Since Cohn (1979, 1980) invented a numerical scheme to solve the orbit-averaged
Fokker-Planck (hereafter FP) equation by direct integration, the scheme has
been widely used as a main tool to study the dynamical evolution of globular
clusters.
Cohn (1979) first performed direct numerical integration of the time-dependent
two-dimensional (hereafter 2D) FP equation in energy-angular momentum $(E,J)$
space with calculation of a self-consistent potential.
This work was done after Cohn and Kulsrud (1978), who studied a steady-state
stellar distribution around a central massive black hole by direct numerical
integration of the 2D FP equation.
Although Cohn (1979) showed the potential power of the direct integration
scheme, he had to stop the calculation at a relatively early stage of
gravothermal core collapse due to numerical error in energy conservation.
The central density of the cluster had increased by only three orders of
magnitude as compared with its initial value, when the calculation was stopped.
Later Cohn (1980) treated one-dimensional (hereafter 1D) energy space FP
equation which was obtained by integrating the $(E,J)$ space FP equation over
$J$-space.
[This $E$-space FP equation was previously employed by H\'enon (1961).]
The averaging over $J$-space means that we assume isotropy of the velocity
distribution, i.e. spherical symmetry in velocity space.

One of the advantages of using the 1D FP equation is that it saves a lot of
computation time and computer memory space.
Another advantage is that numerical error in energy conservation is much
reduced (Cohn 1980).
This reduction of the error is largely owing to the adoption of Chang and
Cooper's (1970) finite-differencing scheme for the 1D FP equation.
The characteristics of the Chang-Cooper scheme are that
it ensures particle conservation and non-negativeness of the distribution
function,
and reproduces the analytic quasi-equilibrium solution exactly.
Due to these characteristics the Chang-Cooper scheme achieves high accuracy
with a relatively small number of meshes.
Cohn (1980) reported that
the secular energy drift rate was reduced by better than a factor of 100
with the adoption of the Chang-Cooper scheme.
In fact he could follow the core collapse until the central density increased
by twenty orders of magnitude.

Because of the above-mentioned advantages of the 1D FP equation,
since the work of Cohn (1980),
the 1D FP equation has been used in most studies on globular cluster evolution
while the 2D FP equation has been seldom used.
The adoption of the 1D $E$-space (isotropic) FP equation seems to be quite
reasonable to study the core evolution which has been a main subject during the
past two decades, because strong relaxation will enforce isotropy of the
velocity distribution in the core.
In fact, Cohn (1985) argued that anisotropy does not significantly alter the
core collapse, although he found that anisotropy extended well inside the
half-mass radius at late times of the core collapse.
Our understanding of the evolution of globular clusters has improved very much
during the past fifteen years, which has been largely owing to the use of the
simple 1D FP equation (cf. Goodman 1993).

The development of anisotropy in the halo is a natural consequence of cluster
evolution driven by two-body relaxation
(Wooley, Robertson 1956; H\'enon 1971; Spitzer, Hart 1971b; Bettwieser, Spurzem
1986).
The relaxation is strong in the core because of its high density and the strong
relaxation continues to produce high-energy stars.
The orbits of the high-energy stars extends into the halo.
Because the density is very low in the halo,
the stars travel the halo almost without experiencing collisions
and return to the core.
Thus the high-energy stars scattered out of the core have very radial orbits on
the average.
Halo stars which are initially not on very radial orbits do not go into the
core and they evolve little.
Therefore radial orbits predominate the halo and velocity anisotropy increases
as the halo grows.

The penetration of anisotropy even into the inner region was pointed out in
several early works (Larson 1970; Spitzer, Shull 1975; Duncan, Shapiro 1982;
Cohn 1985).
Spitzer and Shapiro (1975) argued that it was a natural consequence of the
gravothermal core collapse.
The core becomes hotter as it collapses.
Stars on more nearly radial orbits more reflects this increased temperature
than stars on more nearly circular orbits.
Thus anisotropy in the core is expected to penetrate into the more inner region
as the core collapses.

As described above, the development of velocity anisotropy is expected through
the whole cluster.
So far anisotropy was considered in various simulations of the evolution of
star clusters,
e.g. in calculations of moment equations of the FP equation (Larson 1970;
Bettwieser 1983),
and in Monte-Carlo calculations of the FP equation
(H\'enon 1971; Spitzer, Shull 1975; Duncan, Shapiro 1982).
However, direct 2D FP calculations were carried out only by Cohn(1979, 1985).
Cohn's calculations were for isolated, single-mass clusters without binaries.
Direct 2D FP calculations of more realistic globular cluster models
which include the effects of binaries, stellar mass spectrum, the galactic
tidal field, etc. have not been reported until now.
Therefore how velocity anisotropy changes the evolution of such realistic
cluster models is not known well.

Recently various anisotropic gaseous models and higher-order fluid-dynamical
models of star clusters have been developed
(Bettwieser, Spurzem 1986; Louis 1990; Louis, Spurzem 1991; Spurzem 1991;
Giersz, Spurzem 1994).
These models are composed of moment equations of the FP equation with some
closure relations.
These models have made it possible to calculate in detail the evolution of
anisotropic star clusters.
Some calculations of the post-collapse evolution and of the evolution of
multi-mass clusters have been already done (Giersz, Spurzem 1994; Spurzem 1994;
Spurzem, Takahashi 1995).
However, there are some uncertainties in these models.
It is unclear what closures we should choose, and
whether local description of relaxation effects is really valid for stellar
systems.
Therefore the reliability of anisotropic gaseous and higher-order
fluid-dynamical models should be checked by comparing these results with the
results of more fundamental FP models and/or $N$-body models.
Some such comparisons have been already done (Giersz, Spurzem 1994; Spurzem,
Takahashi 1995);
the results of the anisotropic gaseous models are generally in good agreement
with those of the other models.
However, comparison with the 2D FP models has not been done, because any
reliable 2D FP code was not available.

Now is the good time to reconsider the direct 2D FP calculations.
The sophisticated anisotropic gaseous models are available as described above.
The comparison between the gaseous and the FP models is very useful for
checking the reliability of these models and for interpreting the results of
these models.
On the practical sides of computation, the capability of computers has greatly
developed since the first direct 2D FP calculations were carried out by Cohn
(1979).
At present it is possible to carry out the 2D calculations on standard
workstations .

A main obstacle to the 2D FP calculations is to develop computational schemes
of high accuracy (especially in the energy conservation).
Concerning this problem, a numerical scheme which may be good for 2D FP
calculations was proposed by Takahashi (1993a).
He utilized a scheme based on a generalized variational method instead of the
Chang-Cooper scheme
to solve the FP equation.
Although a classical variational principle for the FP equation does not exist,
a generalized variational principle based on the concept of a local potential
(Glansdorff, Prigogine 1971) does.
Inagaki and Lynden-Bell (1990) gave the local potentials for the orbit-averaged
FP equations.
The generalized variational method was applied by Takahashi and Inagaki (1992)
and Takahashi (1993b) to obtain self-similar solutions of the 1D FP equation.
Takahashi (1993a) solved the time-dependent 1D FP equation by his method
and showed that the method achieved an equal numerical accuracy with the
accuracy achieved by Cohn's method (or the Chang-Cooper scheme).
In Takahashi's method the FP equation is discretized by the finite element
method based on the variational method.
Takahashi (1993a) suggested that his method may be useful for the 2D FP
equation as well, because no special technique applicable only for 1D cases was
used in the method.
On the other hand, any direct 2D generalization of the Chang-Cooper scheme
(which preserves all merits of the Chang-Cooper scheme) has not been found as
far as the author knows.

The aim of the present paper is to develop reliable numerical schemes for the
2D orbit-averaged FP equation and to consider in detail the evolution of
spherical and anisotropic star clusters.
In other words, we improve Cohn's (1979) work so that we can understand the
evolution of anisotropic clusters as well as the evolution of isotropic
clusters.
In the present paper, we consider only the pre-collapse evolution of isolated,
single-mass clusters.
In particular we pay attention to the self-similar structure of the core
collapse and the development of the strongly anisotropic halo.
The effects of binary heating, stellar mass spectrum, the galactic tidal field,
etc. will be considered in future works.

In section \ref{sec:oafp}, the orbit-averaged FP equation in $(E,J)$-space is
described briefly.
In section \ref{sec:method}, the numerical methods to solve the FP equation are
described.
The details of the discretization schemes are given in \ref{sec:ds}.
The accuracy of the present calculations is discussed in section
\ref{sec:accuracy}.
In section \ref{sec:result}, the results of the calculations are presented.
In section \ref{sec:summary}, the results are summarized.

\section{The Orbit-Averaged Fokker-Planck Equation}\label{sec:oafp}

\indent\indent

We study the evolution of spherical one-component star clusters.
We define a distribution function $f({\bf r},{\bf v},t)$
so that $f({\bf r},{\bf v},t)\,d^3{\bf r}\,d^3{\bf v}$ is the number of stars
at time $t$ within the volume element $d^3{\bf r}$ centered at ${\bf r}$ and
within the velocity space element $d^3{\bf v}$ centered at ${\bf v}$.
In a steady-state spherical system, $f$ is a function of only energy $E$ and
angular momentum $J$ per unit mass.
Then the distribution function $f(E,J)$ changes only through collisional
effects.
The time scale of the change is the two-body relaxation time $\tr$ and it is
much longer than the dynamical time scale $t_{\rm d}$ in stellar systems with
large $N$ (cf. Binney, Tremaine 1987; Spitzer 1987).
Thus we can assume that dynamical equilibrium is always established when we
consider the evolution of $f$ caused by collisions.
In this case,
the evolution of $f$ can be described by the orbit-averaged FP equation in
$(E,J)$-space (Cohn 1979).

In numerical calculations, it is more convenient to use the scaled angular
momentum $R$ instead of $J$ as a basic variable.
$R$ is defined by
\begin{equation}
  R=\frac{J^2}{J_{\rm c}^2(E)} \,,
\end{equation}
where $J_{\rm c}(E)$ is the angular momentum of a circular orbit of energy $E$.
Thus $R$ takes all values between 0 and 1, independent of $E$.
($R=0$ corresponds to radial orbits and $R=1$ to circular orbits.)
The number density $N(E,R)$ in $(E,R)$-space is given by
\beqa
  N(E,R) &=& 4\pi^2 P(E,R) J_{\rm c}^2(E) f(E,R) \nonumber \\
         &\equiv& A(E,R)f(E,R) \,,
\eeqa
where $P(E,R)$ is the orbital period.
As Cohn (1979) showed, the 2D FP equation under the fixed gravitational
potential $\phi(r)$ can be written in a flux conserving form
\begin{equation}
A \frac{\partial f}{\partial t} =
- \frac{\partial F_E}{\partial E} - \frac{\partial F_R}{\partial R}
\label{eq:fp} \,,
\end{equation}
where
\begin{eqnarray}
-F_E & = & D_{EE} \frac{\partial f}{\partial E} + D_{ER}\frac{\partial
f}{\partial R} + D_E f \,, \nonumber \\
-F_R & = & D_{RE} \frac{\partial f}{\partial E} + D_{RR}\frac{\partial
f}{\partial R} + D_R f \,.
\end{eqnarray}
The expressions for coefficients $D_{EE}, D_E$, etc. are given in Appendix C of
Cohn (1979).

An isotropized distribution function has been used to derive these expressions
for the coefficients.
The isotropized distribution function $\bar{f}(E,r)$ is defined as
\beq
\bar{f}(E,r) = \frac{1}{2R_{\rm max}^{1/2}} \int_0^{R_{\rm max}}
\frac{dR}{(R_{\rm max}-R)^{1/2}} f(E,R)  \,,
\eeq
where $R_{\rm max}(E,r) = 2r^2[\phi(r)-E]/J_{\rm c}^2(E)$ is the maximum
allowed value of $R$ for all the orbits of energy $E$ which pass through the
radius $r$.
(We adopt the opposite of the usual sign convention for the potential $\phi$
throughout this paper; thus $E=\phi-v^2/2$.)
The isotropization makes the computation of the coefficients much easier.
It has been generally conceived that the use of the isotropized distribution
function does not bring significant inaccuracies, because the coefficients only
depend on moments of $f(E,R)$.
However, as we see below, $f(E,R)$ becomes to strongly depend on $R$
for $E \sim 0$ as the halo develops.
Thus the isotropization may bring significant error in such cases.
On the other hand, since the relaxation occurs virtually only in the core where
the distribution function is almost entirely isotropic, the error may be
negligible.
Although careful studies on the effect of the isotropization is desirable,
we leave such studies to the future.

\section{Methods of Calculation}\label{sec:method}

\indent\indent

The framework of our method is the same as that of Cohn's (1979) method.
Cohn's method is made up of two steps, which are the FP step and the Poisson
step.
In the FP step the distribution function is advanced by solving
the FP equation with the gravitational potential
being held fixed.
In the Poisson step the potential is advanced by solving Poisson's equation
with the distribution function being held fixed as a function of adiabatic
invariants.

A essential difference between our method and Cohn's method exists only in a
discretization scheme of the FP equation.
Here, therefore, we mainly describe procedures in the FP step.

We discretize the 2D FP equation (\ref{eq:fp}) by either the finite difference
method or the finite element method.
Concerning the finite difference method, a Chang-Cooper-like scheme is applied
only to the $E$-direction.
This scheme improves the accuracy of the total energy conservation as compared
with a simple centered difference scheme.
Concerning the finite element method, it was found that the test and weight
functions implied by
the generalized variational principle (Inagaki, Lynden-Bell 1990; Takahashi
1993)
are good for the numerical accuracy.
The details of the discretization schemes are described in \ref{sec:ds}.
After the discretization procedure, the set of linear algebraic equations is
obtained, and it is solved by an appropriate solver of matrix equations.

The FP equation is solved in a rectangular domain enclosed by boundary lines,
$E=\phi(0)$, $E=E_{\rm min}$, $R=0$, and $R=1$,
where $\phi(0)$ is the central potential and the value of $E_{\rm min}$ is
chosen to be close to zero.
We impose boundary conditions that $F_E=0$ on boundaries $E=\phi(0), E_{\rm
min}$
and $F_R=0$ on boundaries $R=0,1$.
The only nontrivial condition is $F_E=0$ on the $E=E_{\rm min}$ boundary.
We choose this condition so that the total mass is conserved.
When we set the value of $E_{\rm min}$ close enough to zero, what boundary
condition we choose may not seriously affect the results we are interested in.

We use variables $(X,Y)$ instead of $(E,R)$ in practical calculations.
The variable $X(E)$ is defined by
\beq
X(E)=\ln\left[\frac{E}{2\phi(0)-E_0-E}\right] \,,
\eeq
where $E_0$ is an adjustable parameter.
This variable was introduced by Cohn (1979).
The value of $E_0$ was chosen to be equal to the energy of a circular orbit at
the core radius.
The variable $Y(R)$ is defined by
\beq
Y(R)=\frac{\ln\left(1+R/R_0\right)}{\ln\left(1+1/R_0\right)} \,,
\eeq
where $R_0$ is an adjustable parameter such as $0 < R_0 \ll 1$.
We set $R_0=0.01$ in standard runs.
This variable is introduced in order to give good representation to radial
orbits.
The necessity of introducing this variable is clear when we see figure 10a.
We set uniform meshes in $X$ and $Y$.
We also need a radial mesh and set a uniform mesh in $\log r$ between $r_{\rm
min}$ and $r_{\rm max}$.
In the standard runs, we chose $r_{\rm min}=10^{-8}r_0$, $r_{\rm max}=100r_0$,
and $E_{\rm min}=\phi(r_{\rm max})$,
where $r_0$ a length scale parameter of Plummer's model chosen as an initial
distribution.

The time step $\Delta t_{\rm P}$ between two succeeding Poisson steps is chosen
in such a way that
the fractional increase in the central density during $\Delta t_{\rm P}$ is
always about 2.5\%.
The time step $\Delta t_{\rm FP}$ used in the FP integration is set to be
$\Delta t_{\rm P}/10$ in the standard runs.

\section{Numerical Accuracy}\label{sec:accuracy}

Numerical error, particularly in energy conservation, was a serious problem in
Cohn's (1979) 2D FP calculations as described in section \ref{sec:intro}.
Due to the problem, direct 2D FP calculations have been seldom performed since
then.
Therefore we examine the numerical accuracy of our calculations before we
discuss the results.

The numbers of grid points in $X$, $Y$, and $r$ are denoted by $N_X$, $N_Y$,
and $N_r$, respectively.
We set $N_X=201$, $N_Y=25$, and $N_r=101$ in the standard runs.
Test runs were carried out with other sets of grid numbers.
The calculations started from Plummer's model were continued until the central
density increased by about 14 orders of magnitude in the standard runs.

Since we used the boundary conditions of vanishing particle-fluxes as described
in section \ref{sec:method}, the total cluster mass should be conserved.
In the FP steps, if we use the finite difference scheme described in appendix
1.1, the total mass is exactly conserved up to round-off error because the flux
conserving finite-difference scheme is used.
However, the mass error is caused by the Poisson steps.
During the calculation the fractional change in the mass was within 0.1\%.
If we use the finite element scheme described in appendix 1.2, the total mass
is not exactly conserved in the FP steps unfortunately.
In this case the fractional change was within 0.3\%.

As Cohn (1979) showed in his Appendix A.II,
the total energy should be conserved, except for the change due to the energy
flux through the $E=E_{\rm min}$ boundary,  by the two-step procedure
consisting of a FP step and a following Poisson step.
The rate of the change of the total (binding) energy ${\cal E}$ (due to the
energy flux through the boundary) is given by
\beq
\frac{d{\cal E}}{dt} =
m \int_0^1 dR \left[EF_E\right]_{E=E_{\rm min}} +
m \int_0^1 dR \left[\frac12 A <(\Delta E)^2>_{\rm t} f \right]_{E=E_{\rm min}},
\label{eq:echg}
\eeq
where $<(\Delta E)^2>_{\rm t}$ is the orbit-averaged diffusion coefficient
concerning the squared change of the energy.
The first term of the right-hand side vanishes because we impose the boundary
condition $F_E(E_{\rm min},R)=0$.
The second term represents a diffusive energy flux through the $E=E_{\rm min}$
boundary.
This term is generally nonzero (positive) unless $f(E_{\rm min},R)=0$,
even if the particle flux $F_E$ is zero at the $E=E_{\rm min}$ boundary.
We might think that this energy change could be negligible since the value of
$f$ was very small at the $E=E_{\rm min}$ boundary.
This is true in 1D $E$-space FP calculations.
In the present 2D calculations, however,
strong anisotropy develops in the halo and the value of $f$ is rather large
near $R=0$ even at the $E=E_{\rm min}$ boundary (see figure 10a).
The amount of the cumulative energy changes given by equation (\ref{eq:echg})
was about 1\% of the total energy at the end of the present calculations.
Considering these changes, we estimated that
the numerical error in the total energy were within 1\% during the
calculations.
In early stages of the calculations, the rate of the energy change
(\ref{eq:echg}) increased rapidly due to the rapid development of anisotropy.
In these stages, the error in the energy mainly came from the FP steps.
The error seemed to be smaller for the finite element code than for the finite
difference code.
In middle stages of the calculations, the energy error was smaller for the
finite difference code when $N_X$ was relatively small (e.g. $N_X$=151).
In late stages of the calculations (then only the core actually evolved), the
energy error mainly came from the Poisson steps and the total (binding) energy
steadily decreased.
We could not definitely conclude which of the two codes was better from a
viewpoint of the energy conservation.
When the simple centered finite-difference scheme was used, the energy error
amounted to about 8\% at the end of the calculation.

The reliability of the calculations can be indirectly tested by comparing the
results obtained by various numerical methods.
In the present paper, the two different discretization schemes were used.
The results obtained by the two schemes were compared in various points as
described in section \ref{sec:result},
and they were generally in good agreement.
This fact supports the reliability of our integration schemes of the FP
equation.
Only noticeable differences between the two schemes appeared in the evolution
of quantities $\xi$ and $\gamma$ defined by equations (\ref{eq:xi}) and
(\ref{eq:gamma}).
These quantities are defined in the core and tends to be constant as the core
collapse proceeds (see figures 4 and 5).
When the finite element scheme was used, the limiting constant values of $\xi$
and $\gamma$ decreased as the number of $X$(energy)-grids $N_X$ increased:
$(\xi\times10^3,\gamma)$=(4.06, 0.111), (3.53, 0.109), (3.28, 0.107) for
$N_X$=101, 151, 201 at the end of each run.
When the finite difference scheme was used, the limiting values varied little
as $N_X$ increased, i.e. the results seemed to have already converged:
$(\xi\times10^3,\gamma)$=(2.98, 0.106), (2.95, 0.105), (2.94, 0.104) for
$N_X$=101, 151, 201 at the end of each run.
The limiting values by the finite element code seem to converge to the limiting
values by the finite difference code as $N_X$ increases.
We may regard the limiting values of $\xi$ and $\gamma$ obtained by the finite
difference code as the real values.

\section{Results}\label{sec:result}

\indent\indent

Calculations were carried out by using both the finite difference and finite
element codes.
The results by the two (partially) different codes were in good agreement
except on the evolution of $\xi$ and $\gamma$ which is described in section
\ref{sec:accuracy}.
Figures shown in this section were actually drawn from the results of the
calculation by the finite difference code.

The initial condition of our calculations is Plummer's model where the velocity
distribution is isotropic everywhere.
We use the standard units of
$G=M=1$ and ${\cal E}_{\rm i}=1/4$, where $G$ is the gravitational, $M$ is the
initial total mass, and ${\cal E}_{\rm i}$ is the initial total (binding)
energy.
No heat sources are included in the present calculations.

Figure 1 shows the evolution of the density profile.
The evolving density profile is plotted every 200 potential-recalculation time
steps.
The central density continues to increase and the core radius continues to
decrease.
This is due to the well-known gravothermal instability (Antonov 1962;
Lynden-Bell, Wood 1968).
The self-similar nature of the gravothermal core collapse (Lynden-Bell,
Eggleton 1980) is clearly seen in figure 1.

The radial profile of the logarithmic density gradient $-\alpha \equiv d\ln
\rho / d\ln r$ is shown in figure 2.
{}From this figure we find that
the inner power-law region of $\rho \propto r^{-2.23}$ develops as the core
collapse proceeds.
This value of $\alpha=2.23$ coincides with that found in a 1D (isotropic) FP
calculation (Cohn 1980).
Heggie and Stevenson (1988) also found the same value of $\alpha=2.23$ for the
pre-collapse self-similar solution of the isotropic FP equation.
A 2D (anisotropic) FP calculation by Cohn (1979) was stopped before the
self-similar structure developed well.

We define an anisotropy parameter $A$ by
\beq
A \equiv 2-2\sigma_{\rm t}^2/\sigma_{\rm r}^2,
\eeq
where $\sigma_{\rm t}$ and $\sigma_{\rm r}$ are the one-dimensional tangential
and radial velocity dispersions, respectively.
[This anisotropy parameter $A$ was used by, e.g.,  Giersz and Spurzem (1994).]
Figure 3 shows the evolving radial profile of $A$.
As the self-similar core collapse proceeds, slight anisotropy proceeds to the
inner region self-similarly.
In the power-law region, $A=0.16$, or $\sigma_{\rm t}^2/\sigma_{\rm r}^2=0.92$.
{}From figure 5 of Cohn (1985) we find the corresponding value of $A \approx
0.15$.
Thus Cohn's calculation is consistent with our calculation in this respect.
However, there are differences between the two calculations with respect to the
development of anisotropy in the outer region of the cluster.
For example, at the half-mass radius and the 90\%-mass radius,
$A$=0.44, 1.62 in our calculation, and
$A$=0.28, 1.24 in Cohn's (1985) calculation,
at late times of the core collapse.
Thus our calculation indicates the development of somewhat stronger anisotropy
in the halo.

It is also useful to compare our results with the results of recent anisotropic
gaseous models.
Louis and Spurzem (1991) obtained pre-collapse (and post-collapse) self-similar
solutions of their anisotropic gaseous models.
Moreover, time-dependent pre-collapse solutions of the anisotropic gaseous
models were shown in figures 2(a) and (b) of Giersz and Spurzem (1994);
these figures can be compared with our figures 2 and 3.
The anisotropic gaseous models include two numerical constants, $\lambda_{\rm
A}$ and $\lambda$ in Giersz and Spurzem's (1994) notation.
The constants $\lambda_{\rm A}$ and $\lambda$ are related to the time-scales of
collisional decay of anisotropy and the heat transport, respectively.
These are free parameters in the context of the gaseous models, and their
values can be determined only through comparison with other models such as
$N$-body models or FP models.
The values of $\alpha$ and $A$ in the gaseous models depend on $\lambda_{\rm
A}\lambda$.
Giersz and Spurzem (1994) plot the values of $\alpha$ and $A$ in the inner
power-law region as a function of $\lambda_{\rm A}\lambda$ in their figures
3(a) and (b).
In these figures, Louis and Spurzem's (1991) self-similar solutions are also
plotted,
and the self-similar solutions and  the time-dependent solutions coincide
rather well.
Giersz and Spurzem (1994) chose $\lambda_{\rm A}=0.1$ and $\lambda=0.4977$ as
standard values through comparison with $N$-body and isotropic FP models.
With these standard values, the gaseous models give $\alpha=2.22$ and $A=0.26$.
This value of $\alpha$ is very close to our value $\alpha=2.23$, and this value
of $A$ is a little higher than but not far from our value $A=0.16$.

Figure 4 presents the evolution of the core collapse rate,
\beq
\xi \equiv t_{\rm r}(0)\frac{d\ln \rho(0)}{d\ln t}, \label{eq:xi}
\eeq
 as a function of the scaled escape energy,
\beq
 x_0 \equiv 3 \phi (0) / v_{\rm m}^2(0),
\eeq
where $t_{\rm r}(0)$ is the central relaxation time and $v_{\rm m}(0)$ is the
central total velocity dispersion.
Here the central relaxation time $t_{\rm r}(0)$ is defined as
\beq
t_{\rm r}(0) \equiv \frac{0.065 v_{\rm m}^3(0)}{G^2 m \rho (0) \ln \Lambda} ,
\eeq
where $m$ is the mass of a star and $\ln \Lambda$ is the gravitational Coulomb
logarithm
(Spitzer, Hart 1971a; Spitzer 1987).
The parameter $\Lambda$ can be expressed as $\Lambda = \mu N$, where $N$ is the
total number of stars in the cluster and $\mu$ is a numerical constant (Spitzer
and Hart adopted $\mu=0.4$).
In the present problem, we do not have to specify the values of $\mu$ and $N$,
because these (in fact, term $N/\ln\Lambda$) can be eliminated from the FP
equation if we use the time measured in units of the relaxation time.
(However, if we include binary heating terms into the FP equation,
we have to specify the values of $\mu$ and $N$.)
The value of $x_0$ monotonically increases as the core collapse proceeds.
Cohn (1980) found that $x_0$ tended to a limiting value of about 13.9,
while we find that $x_0 \approx 13.5$ at the end of our calculation.
If we continued the calculation further, $x_0$ will increase beyond 13.5.
Our calculation gives the asymptotic constant value of $\xi = 2.9 \times
10^{-3}$, while Cohn's (1979) 2D calculation gave $\xi = 6.0 \times 10^{-3}$
and Cohn's (1980) 1D calculation did $\xi = 3.6 \times 10^{-3}$.
Heggie and Stevenson (1988) obtained the value of $\xi = 3.64 \times 10^{-3}$
for the pre-collapse self-similar solution of the 1D FP equation.

While the value of $\xi$ in our anisotropic model is smaller than that in the
isotropic models,
the value of $\xi$ in Cohn's (1979) anisotropic model is larger.
Which tendency is true?
Spitzer (1987, p.95) argued that
the higher value of $\xi$ in anisotropic models was also implied by
a Monte Carlo calculation by Duncan and Shapiro (1982)
Spitzer estimated the value of $\xi \approx 8 \times 10^{-3}$ for that
calculation.
However,
the asymptotic value of $\xi$ in Cohn's (1979) calculation may be less reliable
than that in our calculation,
because Cohn's calculation was accompanied by large error in energy
conservation ($\sim 11\%$) and because the calculation did not follow the core
collapse so deeply.
The reliability of the value given by the Monte Carlo calculation is further
lower.
On the other hand, the lower value of $\xi$ in anisotropic models was argued by
Louis (1990).
Louis obtained the self-similar solutions of isotropic and anisotropic
fluid-dynamical models for star clusters
and found the smaller value of $\xi$ in the anisotropic model.
He discussed the physical reason for the decrease of the core collapse rate in
anisotropic clusters.
He derived a relation among three different types of deviations from thermal
equilibrium, namely the deficiency of high-velocity stars, the temperature
gradient and anisotropy:
the temperature gradient can be balanced by the deficiency of high-velocity
stars which causes a non-zero collision term and thus core collapse, and, for
the anisotropic case, by anisotropy.
Louis argued that generation of anisotropy may be interpreted as an attempt to
avoid core collapse.
Louis's discussion may be expressed in another way as follows.
In anisotropic (i.e. real) clusters, the heat is transferred from the core
more by stars on more radial orbits than by stars on more circular orbits.
In isotropic clusters, however, such a bias to radial orbits cannot occur,
and high-energy circular orbits are enforced to be produced more than in
anisotropic clusters.
High-energy stars on more radial orbits interact with the more inner region
than stars on more circular orbits with the same energy,
or, roughly speaking, the high-energy stars on more radial orbits still (in
part) stay in the core.
This implies that the heat is transferred from the core to the outer region
more quickly in isotropic clusters.
In other words, we may say that the reduction of the 2D FP equation to the 1D
FP equation by averaging over angular momentum space causes artificial
diffusion in addition to real diffusion.
Consequently the core collapse proceeds faster in isotropic clusters.

Figure 5 shows the evolution of the quantity,
\beq
 \gamma \equiv \frac{d\ln v_{\rm m}^2(0)}{d\ln \rho (0)}, \label{eq:gamma}
\eeq
which specifies the equation of state of the core.
The small fluctuations in the curve may be due to numerical error.
Our calculation gives the asymptotic constant value of $\gamma = 0.10$,
while Cohn's (1979, 1980) 2D and 1D calculations gave $\gamma = 0.12, 0.10$,
respectively.
The coincidence of the asymptotic value of $\gamma = 0.10$ between our 2D
calculation and Cohn's (1980) 1D calculation indicates that anisotropy does not
affect the equation of state of the core (in fact, exact isotropy is always
established inside the core).
In a simple homological model of core collapse (Lynden-Bell, Eggleton 1980),
$\gamma$ is related to $\alpha$ as $\gamma=(\alpha-2)/\alpha$.
This relation is well satisfied with the values $\alpha=2.23$ and
$\gamma=0.10$.
The time remaining to complete core collapse, $\tau$, is represented by a fixed
multiple of the current central relaxation time [equation (13) of Cohn (1980)]:
\beq
\tau = [(1-\frac32\gamma)\xi]^{-1} \tr(0).
\eeq
The asymptotic constant values of $\xi$ and $\gamma$ in our calculation gives
$\tau \approx 410\tr(0)$,
while Cohn (1980) obtained $\tau \approx 330 \tr(0)$.

Figure 6a shows the evolution of Lagrangian radii containing inner 1, 2, 5, 10,
20, 30, 40, 50, 75, and 90\% of the cluster mass.
The result of a 1D FP calculation is also shown by dotted curves.
The time is measured in units of the initial half-mass relaxation time $t_{\rm
rh,i}$.
The half-mass relaxation time $t_{\rm rh}$ is defined by
\beq
t_{\rm rh} \equiv 0.138\frac{N^{1/2}r_{\rm h}^{3/2}}{m^{1/2}G^{1/2} \ln
\Lambda},
\eeq
where $r_{\rm h}$ is the radius containing half of the total mass
(Spitzer, Hart 1971a; Spitzer 1987).
We find again slower core collapse in the 2D calculation from figure 6a.
The 2D calculation gives the core collapse time of $t_{\rm coll}=17.6t_{\rm
rh,i}$ and the 1D calculation does $t_{\rm coll}=15.6t_{\rm rh,i}$.
To clarify differences between the 2D and 1D models except in the time scale,
we also present figure 6b where
the time axis of 1D calculation is multiplied by a constant factor
so that the collapse time in the 1D calculation should coincide with that in
the 2D calculation.
In this figure, we do not find any significant differences between the two
models for 1--75\% Lagrangian radii.
However, the 90\% radius of the 2D model expands further than that of the 1D
model, that is, the 2D model has more extended halo.

The density profiles at epochs when the central density increases by about 4.5
orders of magnitude are shown in figure 7 for the 2D model (the solid curve)
and the 1D model (the dotted curve).
We find again that the density profiles for 1D and 2D models are very similar
inside the half-mass radius, but that the 2D model has the higher density in
the outer halo.
The density profile in the outer halo is approximated by a power law $\rho
\propto r^{-3.5}$ rather well for the 2D model,
while no corresponding power law is observed for the 1D model.
(The initial density profile in the halo is asymptotically given by $\rho
\propto r^{-5}$  as indicated by Plummer's density law.)
This power law is not strictly established, but the value of the logarithmic
density gradient is between $-3$ and $-4$ in the halo (figure 2).
Spitzer and Shapiro (1972) demonstrated by using the $(E,J)$-space FP equation
that the steady state density profile in the halo is given by $\rho \propto
r^{-3.5}$.
Spitzer and Shull (1975) showed this power law was observed in their
Monte-Carlo calculation started from Plummer's model.
However, Cohn (1979) did not observe this power law.
He stated this discrepancy might be due to the use of an energy cutoff in his
calculation, corresponding to a tidal limiting sphere of radius $r_{\rm
t}=100r_0$ ($r_0$ is a length scale parameter of Plummer's model); the density
must vanish at $r=r_{\rm t}$ in his calculation, so that $d\ln\rho/d\ln r$ must
drop considerably near $r=r_{\rm t}$.
In the present calculation we used different boundary conditions: the flux
$F_E=0$ on an energy cutoff boundary $E=E_{\rm min}$.
Therefore the density does not have to vanish at the outermost radial grid
point.

The well-developed halo in the 2D model is due to the emergence of stars on
very eccentric orbits from the core into the halo.
The halo is dominated by these eccentric orbits.
This is clearly seen in figure 8 which shows the radial and tangential velocity
dispersion profiles for the 2D model at the same epoch as in figure 7.
In the halo the radial velocity dispersion exceeds the tangential velocity
dispersion considerably.
We find from figure 3 that the value of the anisotropy parameter $A$ increases
monotonically and approaches to the maximum value of 2 as the radius increases.
A power law profile $\sigt^2 \propto r^{-2}$ in the outer halo, which
corresponds to the constant mean square angular momentum,  were indicated by
several authors (e.g. Spitzer, Hart 1971b; H\'enon 1971).
We find from figure 8 that this power law gives a reasonable fit to the result
of the 2D FP calculation.
(The initial velocity dispersion profile in the halo is asymptotically given by
$\rho \propto r^{-1}$.)

Figure 9 shows the evolution of the mass-averaged velocity dispersions
between 75\% and 90\% Lagrangian radii.
Giersz and Heggie (1994a) showed the same figure which was produced from their
statistical data of 1000-body calculations.
The evolution of the velocity dispersions in our model seems to be
qualitatively similar to that in the 1000-body model.
However, there are some quantitative differences: e.g., in the 1000-body model
$\sigr^2 \approx 2\sigt^2 \approx 0.075$ at the collapse time, while in our 2D
FP model $\sigr^2 \approx 0.055, 2\sigt^2 \approx 0.04$ at the collapse time.
Therefore the value of the anisotropy parameter $A$ is higher in the 2D model.
We note the core collapse was not actually completed in the 1000-body model;
the core contraction stopped due to binary heating when the central density
increased only by about two orders of magnitude.
The lower value of $A$ in the 1000-body model may be due to this incomplete
core collapse.

In figure 10a, the distribution function $f$ at the same epoch as in figure 7
is plotted as a function of the scaled angular momentum $R$.
Different curves correspond to different energies (see the figure caption):
the upper curves correspond to larger $E$ (lower energy in a usual sense) and
the lower curves to smaller $E$ (higher energy).
For large $E$, the distribution function $f$ is actually independent of the
angular momentum $R$, and is very close to an isotropic Maxwellian one.
For small $E$, $f$ increases very rapidly as $R$ goes to zero,
and $f$ for larger $R$ has been almost unchanged from its initial value.
As described above, the halo is predominated by stars of very eccentric orbits
($R \approx 0$).
These stars come from the inner region of the cluster.
This is evident from figure 10b
where $f$ is plotted as in figure 10a, but the abscissa is the pericenter
radius $r_{\rm p}(E,R)$.
For small $E$, $f$ has large values when $r_{\rm p}$ lies within the inner
region of the cluster (the initial half-mass radius is $r_{\rm h,i} \approx
0.77$).

In figures 6 and 9, the time is measured in units of the initial half-mass
relaxation time.
To obtain the physical time, we must specify the number of stars $N$ and the
value of the coefficient $\mu$ in the Coulomb logarithm $\ln(\mu N)$.
While a conservative value is $\mu=0.4$ (Spitzer, Hart 1971a),
Giersz and Heggie (1994a) obtained $\mu=0.11$ by comparing $N$-body models with
isotropic gaseous and isotropic FP models.
Various estimates of $\mu$ by other authors are listed in table 2 of Giersz and
Heggie (1994a).
Rough comparison between our 2D FP calculation and Giersz and Heggie's (1994a)
1000-body calculation with respect to the evolution of the Lagrangian radii
indicates that the value of $\mu=0.11$ gives a reasonably good agreement of the
two results.
Since the core collapse proceeds more slowly in the 2D model than in the 1D
model as described above, the value of $\mu$ somewhat larger than $0.11$ may
give a better agreement.
However, comparison with $N$-body models is complicated by the inevitable
formation of binaries in $N$-body models;
 the core collapse stops due to the binary heating at relatively early stages
in small $N$-body systems.
Detailed comparison as Giersz and Heggie (1994a, b) did is necessary to
determine the best empirical value of $\mu$.
The initial half-mass relaxation time is $t_{\rm rh,i}=19.67$ for $\mu=0.11$
and $N=1000$.

\section{Summary}\label{sec:summary}

\indent\indent

We have developed the numerical codes to solve the orbit-averaged $(E,J)$-space
FP equation with high accuracy and have investigated in detail the pre-collapse
evolution of single-mass spherical star clusters where velocity anisotropy is
allowed.
The purpose of this work is to improve Cohn's (1979) work in order to
understand the evolution of {\it anisotropic} star clusters thoroughly.
Anisotropic (2D) FP models have been seldom used for the last fifteen years
because of the difficulties of the numerical calculations.
In particular the accuracy of total energy conservation was bad in 2D FP
calculations.
It was thought that the origin of this error was in the integration scheme of
the 2D FP equation.

In this paper we have developed two different integration schemes.
One is the finite difference scheme where the Chang-Cooper scheme is simply
applied only for the energy direction.
The other is the finite element scheme where the test and weight functions
suggested by the generalized variational principle are used.
Using these schemes we could follow the core collapse
until the central density increased by about 14 orders of magnitude.
The total energy error was about 1\% at the ends of these calculations.
We believe that the accuracy of the present calculations is enough to trust the
quantitative results presented in section \ref{sec:result}.

The results obtained by the finite difference and the finite element schemes
are generally in good agreement.
Although the finite difference scheme seemed to be better to follow the core
evolution at late stages of the core collapse,
we could not definitely conclude which of the two schemes was better through
the whole evolutionary stages.
The accuracy of the finite element scheme may be improved by using higher-order
basis functions.
The sources of the numerical errors are not only in the integration of the FP
equation but also in the calculation of the diffusion coefficients and in the
potential-recalculation steps.
In these calculations many multi-dimensional numerical integrations and
interpolations are involved, which are of course more difficult to perform than
one-dimensional ones.
If these calculation schemes are improved, the numerical errors will be reduced
further.

Self-similar evolution of the core collapse is clearly seen in the present
calculations.
In the self-similar region between the isothermal core and the outer halo, the
density profile is characterized by a power law $\rho \propto r^{-2.23}$ and
velocity anisotropy is characterized by a constant ratio $\sigma_{\rm
t}^2/\sigma_{\rm r}^2=0.92$.
This power-law index of the density profile is the same as that in the
isotropic FP model.
As the core collapse proceeds, the core radius and mass go to zero and the
velocity anisotropy penetrates into the inner region.
Therefore, when the core collapse completes, the velocity distribution is
anisotropic everywhere in the cluster.
At late stages of the core collapse,
the collapse rate $\xi \equiv t_{\rm r}(0)d\ln \rho(0)/d\ln t$ and
the quantity $\gamma \equiv d\ln v_{\rm m}^2(0)/d\ln \rho (0)$
tend to asymptotic constant values $\xi=2.9 \times 10^{-3}$ and $\gamma=0.10$.
The asymptotic values in the isotropic model are $\xi=3.6 \times 10^{-3}$ and
$\gamma=0.10$.
With these values of $\xi$ and $\gamma$, we can estimate that
the time remaining until complete collapse, $\tau$, is $\tau=410\tr(0)$ in the
anisotropic model, and $\tau=330\tr(0)$ in the isotropic model.
The collapse time is $17.6t_{\rm rh,i}$ in the anisotropic model and
$15.6t_{\rm rh,i}$ in the isotropic model, when Plummer's model is chosen as
the initial condition.

In the outer half of the mass, strong velocity anisotropy has developed.
At late times the ratio of the tangential velocity dispersion to the radial one
decreases monotonically to nearly zero as the radius increases:
e.g., $\sigt^2/\sigr^2 \approx 0.78, 0.19$ at the half-mass and the 90\%-mass
radii, respectively, at the end of the collapse.
The tangential velocity dispersion profile in the outer halo is reasonably
approximated by a power law $\sigt^2 \propto r^{-2}$.
This indicates that the mean square angular momentum is almost constant
independent of the radius.
Although the density profiles for 1D and 2D models are very similar inside the
half-mass radius, the 2D model has the higher density in the outer halo.
The density profile in the outer halo is approximated by a power law $\rho
\propto r^{-3.5}$, which was indicated by Spitzer and Shapiro (1972).
The well-developed halo in the 2D model is dominated by eccentric orbits.
These eccentric orbits have been produced by strong relaxation in the core.

In this paper the detailed direct 2D FP calculations have been performed.
We have seen that the relaxation process is always accompanied by the velocity
anisotropy production.
We have found that the core collapse in single-mass clusters is not seriously
affected by anisotropy,
though the core collapse rate is reduced a little by the development of the
anisotropy.
However, we do not know very much the effects of anisotropy on the
post-collapse evolution, and on the evolution of more realistic clusters (e.g.,
multi-mass clusters, tidally limited clusters, etc.).
The studies of the effects by using anisotropic gaseous models have been
progressing (Giersz, Spurzem 1994; Spurzem 1994; Spurzem, Takahashi 1994).
On the other hand, the 2D FP models have been seldom used during the last
fifteen years, mainly due to the numerical difficulties.
The present paper has shown that 2D FP calculations can be performed with
reasonable numerical accuracy.
We believe that we do not have to hesitate to do 2D FP calculations now.
The evolution of more realistic cluster models will be studied in future works.

\bigskip\bigskip
I would like to thank Prof. S. Inagaki for valuable advice and for giving me
his 1D FP code which served as a good guide to the making of my 2D FP codes.
I thank Dr. M. H. Lee for kindly showing me his finite-differencing scheme used
in his 2D FP code for rotating star clusters.
That was very helpful for my developing the present finite-differencing scheme.
I also thank Dr. M. Itoh for giving me his 2D FP solver which is a original of
the present finite-difference scheme.
The early part of this work was carried out during my stay at the Institute for
Theoretical Physics, Santa Barbara, California, USA, supported by the NSF under
Grant No. PHY89-04035.
This work was supported in part by the Grand-in-Aid for Encouragement of Young
Scientists by the Ministry of Education, Science and Culture of Japan (No.
1338).

\appendix

\renewcommand{\thesection}{Appendix\ \arabic{section}}
\renewcommand{\thesubsection}{Appendix \arabic{section}. \arabic{subsection}}
\renewcommand{\theequation}{A\arabic{equation}}
\setcounter{equation}{0}

\section{Discretization Schemes} \label{sec:ds}

\indent\indent

In this appendix we describe the discretization schemes of the FP equation
(\ref{eq:fp}) in detail.
Two independent variables are denoted by $(x,y)$ here.
The discretization is done by using the finite difference method and the finite
element method.

\subsection{The Finite Difference Scheme}

\indent\indent

We represent $f(x,y,t)$ by the discrete set $f_{i,j}^n \equiv f(x_i,y_j,t_n)$.
We write $\Delta x$, $\Delta y$, and $\Delta t$
for $x$, $y$, and time mesh sizes, respectively:
$x_{i+1} \equiv x_i+\Delta x$, $y_{j+1} \equiv y_j+\Delta y$, and $t_{n+1}
\equiv t_n+\Delta t$.
We do the finite differencing of the FP equation as
\beq
A_{i,j}^n \frac{f_{i,j}^{n+1}-f_{i,j}^n}{\Delta t} =
-\frac{\tilde{F}_{x\, i+\hf,j}-\tilde{F}_{x\, i-\hf,j}}{\Delta x}
-\frac{\tilde{F}_{y\, i,j+\hf}-\tilde{F}_{x\, i,j-\hf}}{\Delta y},
\eeq
where
\beqa
\tilde{F}_{x\, i+\hf,j} &=& -D_{x\, i+\hf,j}^n \tilde{f}_{i+\hf,j}
-D_{xx\, i+\hf,j}^n \frac{\tilde{f}_{i+1,j}-\tilde{f}_{i,j}}{\Delta x}
-D_{xy\, i+\hf,j}^n
\frac{\tilde{f}_{i+\hf,j+\hf}-\tilde{f}_{i+\hf,j-\hf}}{\Delta y},
\label{eq:fdfluxx} \\
\tilde{F}_{y\, i,j+\hf} &=& -D_{y\, i,j+\hf}^n \tilde{f}_{i,j+\hf}
-D_{yx\, i,j+\hf}^n
\frac{\tilde{f}_{i+\hf,j+\hf}-\tilde{f}_{i-\hf,j+\hf}}{\Delta x}
-D_{yy\, i,j+\hf}^n \frac{\tilde{f}_{i,j+1}-\tilde{f}_{i,j}}{\Delta y},
\label{eq:fdfluxy}
\eeqa
and similar expressions for the other flux terms.
Here, the diffusion coefficients are evaluated at $t=t_n$,
and $\tilde{f_j}$ is some linear combination of $f_j^n$ and $f_j^{n+1}$.
In this paper, we adopt the Crank-Nicolson scheme: $\tilde{f_j} =
(f_j^n+f_j^{n+1})/2$.
Hereafter we omit the tildes.
In the right side of equation (\ref{eq:fdfluxx}), $f_{i+\hf,j+\hf}$ is given by
\beq
f_{i+\hf,j+\hf}=\frac14 (f_{i,j}+f_{i+1,j}+f_{i,j+1}+f_{i+1,j+1}),
\eeq
and $f_{i+\hf,j-\hf}$ is given by a similar expression.
We represent $f_{i+\hf,j}$ in the first term of equation (\ref{eq:fdfluxx}) as
\beq
f_{i+\hf,j}=\delta_{x\,i,j}f_{i,j}
+(1-\delta_{x\,i,j})f_{i+1,j} .
\eeq
As in the Chang-Cooper scheme, a key point is a choice of the value of
parameter $\delta_{x\,i,j}$.
We may simply choose $\delta_{x\,i,j}=1/2$.
However, imitating the Chang-Cooper scheme, we determine its value as follows.
First we consider the equilibrium solution of the FP equation.
If we assume that $f$ is independent of $y$ and that the diffusion coefficients
are independent of $f$,
the solution of $F_x=0$ is
\beq
f= C \exp \left[-\int \frac{D_x}{D_{xx}} dx\right],
\eeq
where $C$ is a constant of integration.
Then, in the equilibrium state,
\beqa
\frac{f_{i+1,j}}{f_{i,j}}&=&
\exp\left[-\int_{x_i}^{x_{i+1}}\frac{D_x(x',y_j,t)}{D_{xx}(x',y_j,t)}dx'\right]
 \nonumber \\
&\approx& \exp\left[-\frac{D_{x\, i+\hf,j}}{D_{xx\, i+\hf,j}}\Delta x\right].
\eeqa
We choose $\delta_{x\,i,j}$ so that this relation will be reproduced
by setting the finite-differenced form of the flux (\ref{eq:fdfluxx}) to be
zero with the neglect of the third term.
Then we find
\beq
\delta_{x\,i,j} = \frac{1}{w_{x\,i,j}} - \frac{1}{\exp(w_{x\,i,j})-1},
\eeq
where
\beq
w_{x\,i,j}=\Delta x \frac{D_{x\, i+\hf,j}}{D_{xx\, i+\hf,j}}.
\eeq
The parameter $\delta$ monotonically decreases from 1 to 0 as $w$ goes from
$-\infty$ to $+\infty$, and $\delta=0.5$ for $w=0$.
Thus the value of $\delta$ is significantly different from 0.5 where the
advection dominates over the diffusion, i.e. $|w_x|$ is large.
We can take the same procedure for the finite-differencing of $F_y$.

This scheme is not a real 2D generalization of the Chang-Cooper scheme.
In fact, it does not ensure non-negativeness of the distribution function,
nor does it preserve the general equilibrium solution.
However, if the problem is reduced to the 1D problem (i.e. $f$ depends only on
$x$ or $y$) in some region,
the present scheme is virtually reduced to the (1D) Chang-Cooper scheme.
In the present problem, since an isotropic Maxwellian distribution is almost
established in the core, our scheme is expected to work well there.
In fact the adoption of our scheme reduced the total energy error by a factor
of more than ten relative to the simple centered differencing.
Numerical experiments revealed that this Chang-Cooper--like differencing for
$R$-direction did not affect the results, and that $\delta_R$ was always close
to 0.5 everywhere.
Therefore we simply set $\delta_R=0.5$ in the standard runs.

Cohn (1985) also reported that he investigated several alternative
generalizations of the Chang-Cooper scheme and that all of these improved
energy conservation.
Although his schemes were not explained in detail, our scheme is possibly
similar to them.

\subsection{The Finite Element Scheme}

\indent\indent

We approximate the solution $f$ by using a set of basis functions
$\{\varphi_1,\varphi_2,\dots,\varphi_N\}$ as
\beq
g(x,y,t) \equiv \log f(x,y,t) = \sum_{j=1}^N g_j(t) \varphi_j(x,y)
\label{eq:femg} \,.
\eeq
Applying the method of weighted residuals, we multiply equation (\ref{eq:fp})
by weight $\varphi_i$ and integrate it over the entire domain $\Omega$.
Thus we obtain
\beq
\int\!\!\int_\Omega A \frac{\pa f}{\pa t} \varphi_i \,dx\,dy
=-\int\!\!\int_\Omega \left( \frac{\pa F_x}{\pa x} + \frac{\pa F_y}{\pa y}
\right)\varphi_i \,dx\,dy \,.
\eeq
The right side of this equation can be rewritten by application of Green's
theorem, which produces
\beq
\int\!\!\int_\Omega A \frac{\pa f}{\pa t} \varphi_i \,dx\,dy
= \int\!\!\int_\Omega \left( F_x \frac{\pa \varphi_i}{\pa x} + F_y \frac{\pa
\varphi_i}{\pa y} \right) \,dx\,dy
-\int_\Gamma \varphi_i {\bf F}\cdot{\bf n} \,d\Gamma \label{eq:wrm} \,,
\eeq
where ${\bf n}$ denotes the unit normal to the boundary $\Gamma$ of $\Omega$,
directed outward.
The second term of the right side of this equation vanishes if $\varphi_i
\equiv 0$ or ${\bf F}\cdot{\bf n} \equiv 0$ on the boundary $\Gamma$.
Because this condition holds true in the present case, we omit this term
hereafter.
Substituting equation (\ref{eq:femg}) into equation (\ref{eq:wrm}), we obtain
\beq
\sum_j m_{ij}\frac{dg_j}{dt} = \sum_j k_{ij}g_j+l_i \,, \label{eq:fem0}
\eeq
where
\beqa
m_{ij}&=&\int\!\!\int_\Omega Af\varphi_j\varphi_i \,dx\,dy , \nonumber \\
k_{ij}&=& -\int\!\!\int_\Omega \left[ \left( D_{xx}\frac{\pa\varphi_j}{\pa x} +
D_{xy}\frac{\pa\varphi_j}{\pa y} \right) \frac{\pa\varphi_i}{\pa x}
+ \left( D_{yx}\frac{\pa\varphi_j}{\pa x} + D_{yy}\frac{\pa\varphi_j}{\pa y}
\right) \frac{\pa\varphi_i}{\pa y} \right] f \,dx\,dy , \nonumber \\
l_{i}&=&\int\!\!\int_\Omega \left( D_x\frac{\pa\varphi_i}{\pa x} +
D_y\frac{\pa\varphi_i}{\pa y} \right) f \,dx\,dy .
\eeqa
Then equation (\ref{eq:fem0}) is discretized with respect to time as
\beq
\sum_j m_{ij}^n \frac{g_j^{n+1}-g_j^n}{\Delta t}
=\sum_j k_{ij}^n \tilde{g_j} + l_i^n \,, \label{eq:fem1}
\eeq
where $\tilde{g_j}$ is a linear combination of $g_j^n$ and $g_j^{n+1}$.
In fact we take $\tilde{g_j} = (g_j^{n+1}+g_j^n)/2$.
We simply evaluate $m_{ij}$, $k_{ij}$, and $l_i$ at $t=t_n$.

We use rectangular elements to discretize the domain $\Omega$ .
Node points are located at the four corners of the element.
Two-dimensional piecewise bilinear polynomials are used as the basis functions.
Concerning the more detailed procedure of the finite element method, see
appropriate textbooks (e.g., Fletcher 1991).

We may expand solution $f$ as $f=\sum f_j(t)\varphi_j(x,y)$ instead of equation
(\ref{eq:femg}).
Test calculations by using the finite element scheme with this trial function
were also carried out.
The mass and energy errors were larger than in the calculation with the trial
function (\ref{eq:femg}).
The value of $f$ varies by many orders of magnitude over the domain where the
equation is solved.
This indicates that we had better use $\ln f$ as a variable.
This may be one simple reason why the trial function (\ref{eq:femg}) is better.
We also note that the form of the trial function (\ref{eq:femg}) is naturally
derived from Inagaki and Lynden-Bell's (1990) generalized variational
principle.
Taking the variation of the local potential $\Phi$ [equation (5.5) of Inagaki
and Lynden-Bell (1990)], we obtain
\beqa
\delta \Phi
 &=& \int\!\!\int_\Omega A \frac{\pa f}{\pa t} \delta \ln f \,dx\,dy
-\int\!\!\int_\Omega \left( F_x \frac{\pa \delta \ln f}{\pa x}
   +F_y \frac{\pa \delta \ln f}{\pa y} \right) \,dx\,dy \nonumber \\
&=&0 .
\eeqa
When we adopt the trial function (\ref{eq:femg}), we obtain equation
(\ref{eq:wrm}) again.


\begin{reference}
\item Antonov V.A. 1962, Vestn. Leningrad Gros. Univ. 7, 135
[translated in Dynamics of Star Clusters, IAU Symp No. 113, 1985, ed. J.
Goodman, P. Hut (D.~Reidel Publishing Company, Dordrecht) p525].
\item Bettwieser E. 1983, \mn\ 203, 811.
\item Bettwieser E., Spurzem R. 1986, \aap\ 161, 102.
\item Binney J., Tremaine S. 1987, Galactic Dynamics (Princeton University
Press, Princeton).
\item Chang J.S., Cooper G. 1970, J. Comp. Phys. 6, 1.
\item Cohn H. 1979, \apj\ 234, 1036.
\item Cohn H. 1980, \apj\ 242, 765.
\item Cohn H. 1985, in Dynamics of Star Clusters, IAU Symp No. 113, ed. J.
Goodman , P. Hut (D.~Reidel Publishing Company, Dordrecht) p161.
\item Cohn H., Kulsrud R.M. 1978, \apj\ 226, 1087.
\item Duncan M.J., Shapiro S.L. 1982, \apj\ 253, 921.
\item Fletcher C.A.J., 1991, Computational Techniques for Fluid Dynamics Vol. I
(Springer-Verlag, Berlin) ch5.
\item Giersz M., Heggie D.C. 1994a, \mn\ 268, 257.
\item Giersz M., Heggie D.C. 1994b, \mn\ 270, 298.
\item Giersz M., Spurzem R. 1994, \mn\ 269, 241.
\item Glansdorff P., Prigogine I. 1971, Thermodynamic Theory of Structure,
Stability and Fluctuations (Wiley-Interscience, London) chX.
\item Goodman J. 1993, in Structure and Dynamics of Globular Clusters, ed. S.G.
Djorgovski, G. Meylan (Astron. Soc. Pac., San Francisco) p87.
\item Heggie D.C., Stevenson D. 1988, \mn\ 230, 223.
\item H\'enon M. 1961, Ann. Astrophys. 24, 369.
\item H\'enon M. 1971, ApSS 13, 284.
\item Inagaki S., Lynden-Bell D. 1990, \mn\ 244, 254.
\item Larson R.B. 1970, \mn\ 150, 93.
\item Louis P.D. 1990, \mn\ 244, 478.
\item Louis P.D., Spurzem R. 1991, \mn\ 244, 478.
\item Lynden-Bell D., Eggleton P.P. 1989, \mn\ 191, 483.
\item Lynden-Bell D., Wood R. 1968, \mn\ 138, 495.
\item Spitzer L. Jr. 1987, Dynamical Evolution of Globular Clusters (Princeton
University Press, Princeton).
\item Spitzer L. Jr., Hart M.H. 1971a, \apj\ 164, 399.
\item Spitzer L. Jr., Hart M.H. 1971b, \apj\ 166, 483.
\item Spitzer L. Jr., Shapiro S.L. 1972, \apj\ 173, 529.
\item Spitzer L. Jr., Shull J.M. 1975, \apj\ 200, 339.
\item Spurzem R. 1991, \mn\ 252, 177.
\item Spurzem R. 1994, in Ergodic Concepts in Stellar Dynamics, ed.  D.
Pfenniger, V.A. Gurzadyan (Springer, Berlin) p170.
\item Spurzem R., Takahashi K. 1995, \mn\ 272, 772.
\item Takahashi K. 1993a, \pasj\ 45, 233.
\item Takahashi K. 1993b, \pasj\ 45, 789.
\item Takahashi K., Inagaki S. 1992, \pasj\ 44, 623.
\item Wooley R.v.d.R., Robertson D.A. 1956, \mn\ 116, 288.
\end{reference}

\clearpage

\leftline{\Large\bf Figure Captions}

\bigskip

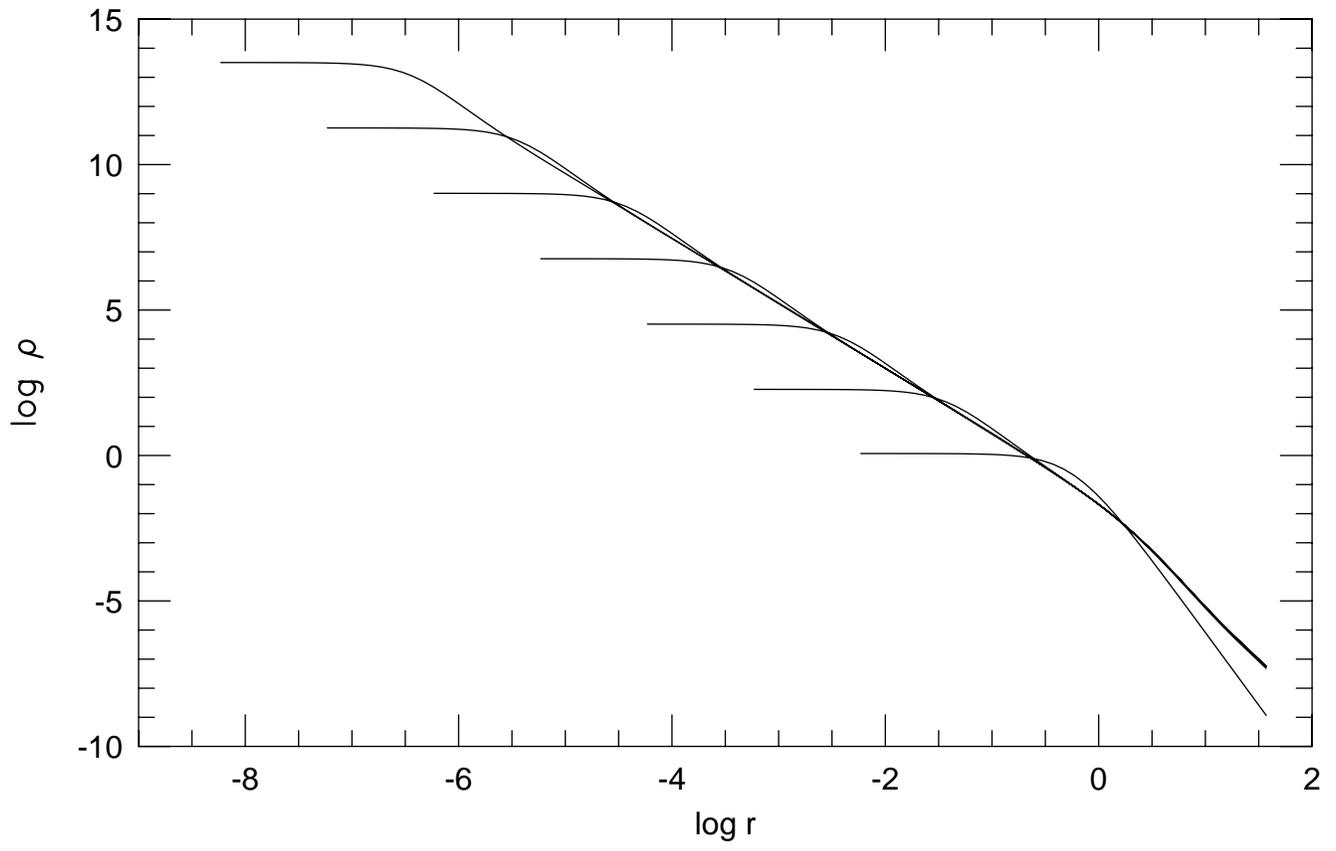
{\bf Fig. 1.}
Evolution of the density profile.
The evolving density profile is plotted every 200 potential-recalculation time
steps.
The central density increases and the core radius decreases with time.

\medskip

{\bf Fig. 2.}
Evolution of the radial profile of the logarithmic density gradient $d\ln \rho
/ d\ln r = -\alpha$.
The profiles at the same epochs as in figure 1 are plotted.
The power-law region with $\alpha=2.23$ extends into the inner region as the
core collapse proceeds.

\medskip

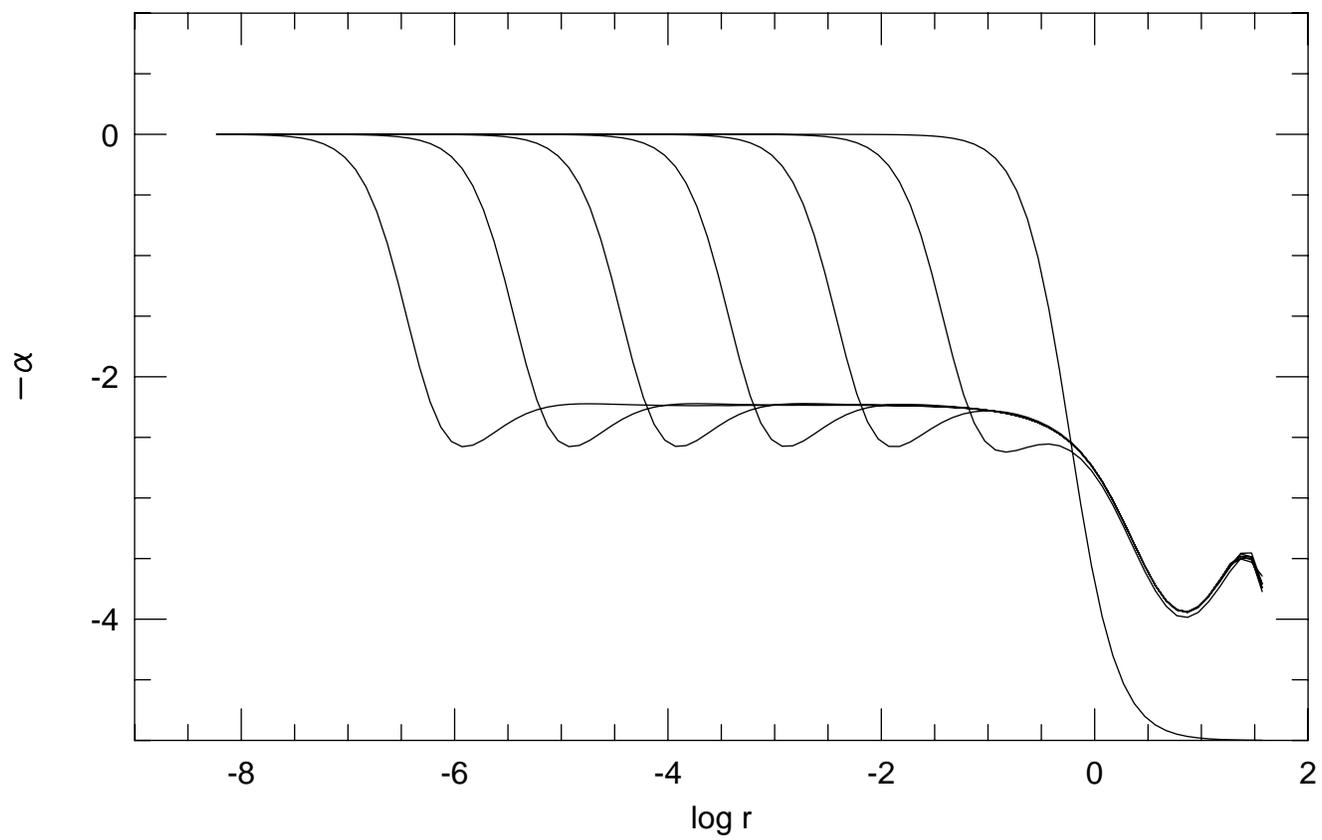
{\bf Fig. 3.}
Evolution of the radial profile of the anisotropy parameter
$A \equiv 2-2\sigma_{\rm t}^2/\sigma_{\rm r}^2$.
The profiles at the same epochs as in figure 1 are plotted.
The self-similar region with $A=0.16$ extends into the inner region as the core
collapse proceeds.

\medskip

{\bf Fig. 4.}
Evolution of the core collapse rate
$\xi \equiv t_{\rm r}(0)[d\ln \rho(0)/d\ln t]$
as a function of the scaled escape energy $x_0 \equiv 3\phi(0)/v_{\rm m}^2(0)$.
The asymptotic constant value of $\xi$ is $\xi = 2.9 \times 10^{-3}$.

\medskip

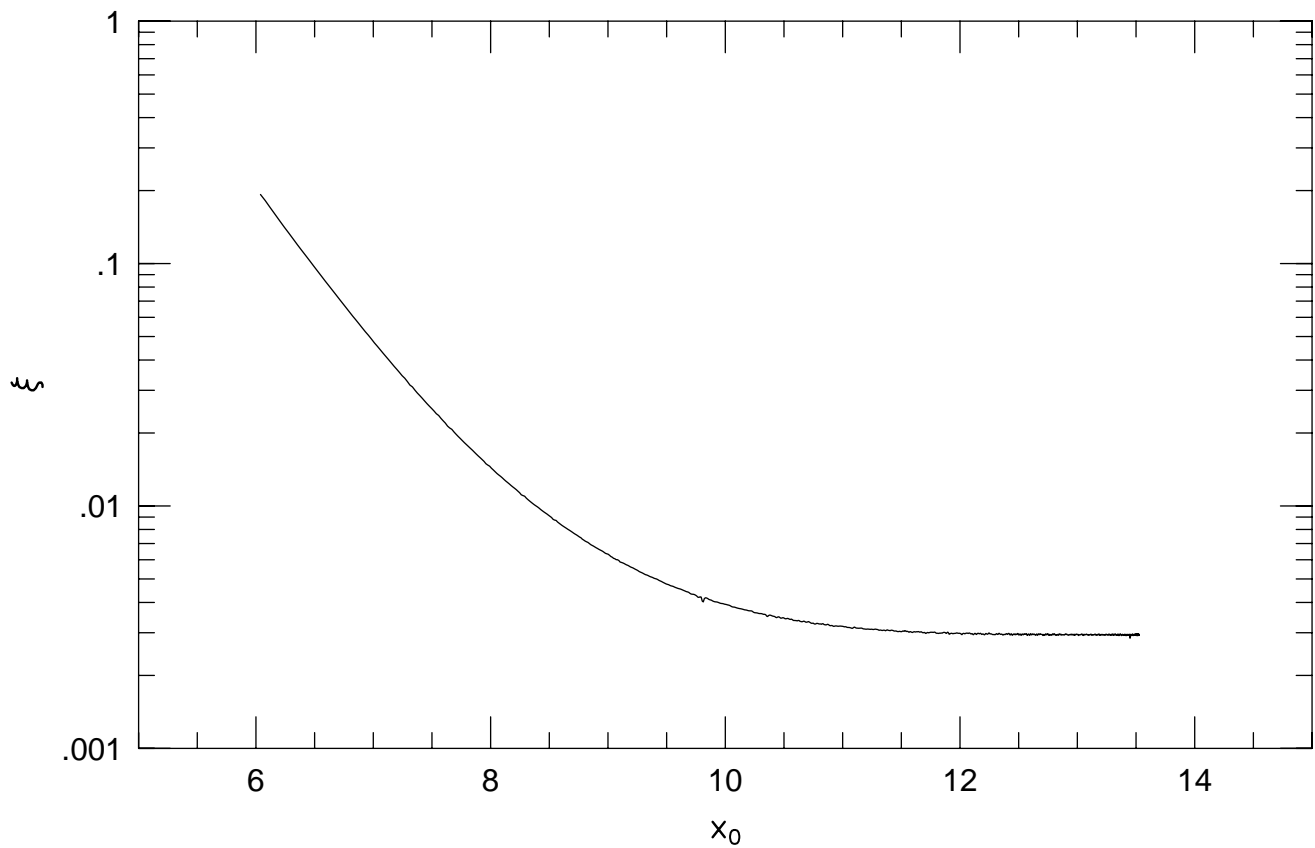
{\bf Fig. 5.}
Evolution of the quantity $\gamma \equiv d\ln v_{\rm m}^2(0)/d\ln \rho(0)$ as a
function of $x_0$.
The asymptotic constant value of $\gamma$ is $\gamma = 0.10$.
The small fluctuations in the curve may be due to numerical error.

\medskip

{\bf Fig. 6.}
(a) Evolution of Lagrangian radii containing inner 1, 2, 5, 10, 20, 30, 40, 50,
75, and 90\% of the cluster mass.
The solid curves are the result of the 2D FP calculation, while the dotted
curves are that of 1D calculation.
The time is measured in units of the initial half-mass relaxation time $t_{\rm
rh,i}$.
(b) Same as (a), but
the time axis of 1D calculation is multiplied by a constant factor
so that the collapse time in the 1D calculation should coincide with that in
the 2D calculation.

\medskip

{\bf Fig. 7.}
The density profile at the 400th potential-recalculation time step
($t=17.5t_{\rm rh,i}$) in the 2D FP calculation is shown by the solid curve.
The dotted curve is the density profile in the 1D calculation at an epoch when
the central density is about the same as in the 2D calculation ($t=15.6t_{\rm
rh,i}$).
The asymptotic line $\rho \propto r^{-3.5}$ is shown for comparison.

\medskip

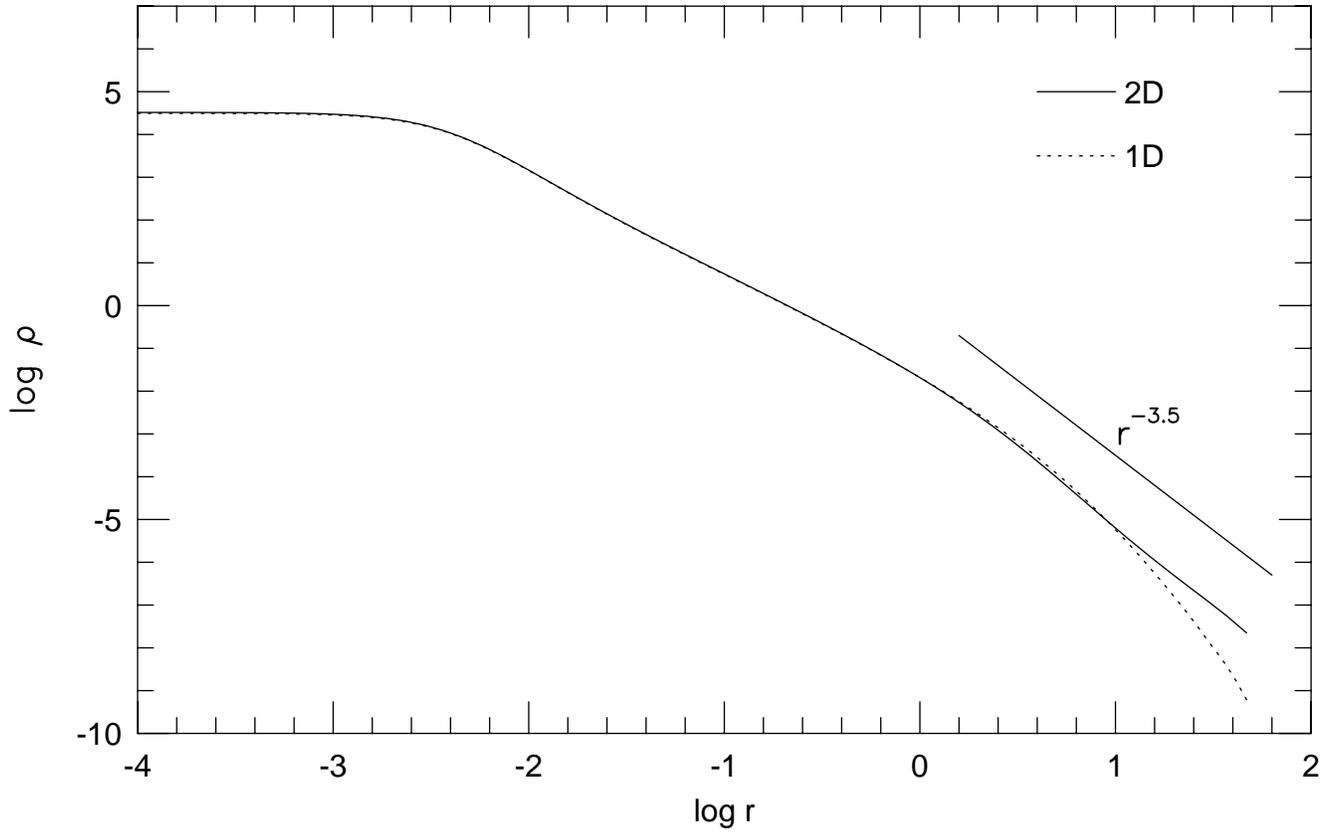
{\bf Fig. 8.}
The velocity dispersion profiles at the same epoch as in figure 7.
The solid and dashed curves are the radial and 1D tangential velocity
dispersions in the 2D FP calculation, respectively.
The dotted curve is the 1D velocity dispersion profile in the 1D FP
calculation.
The asymptotic line $\sigma^2 \propto r^{-2}$ is shown for comparison.

\medskip

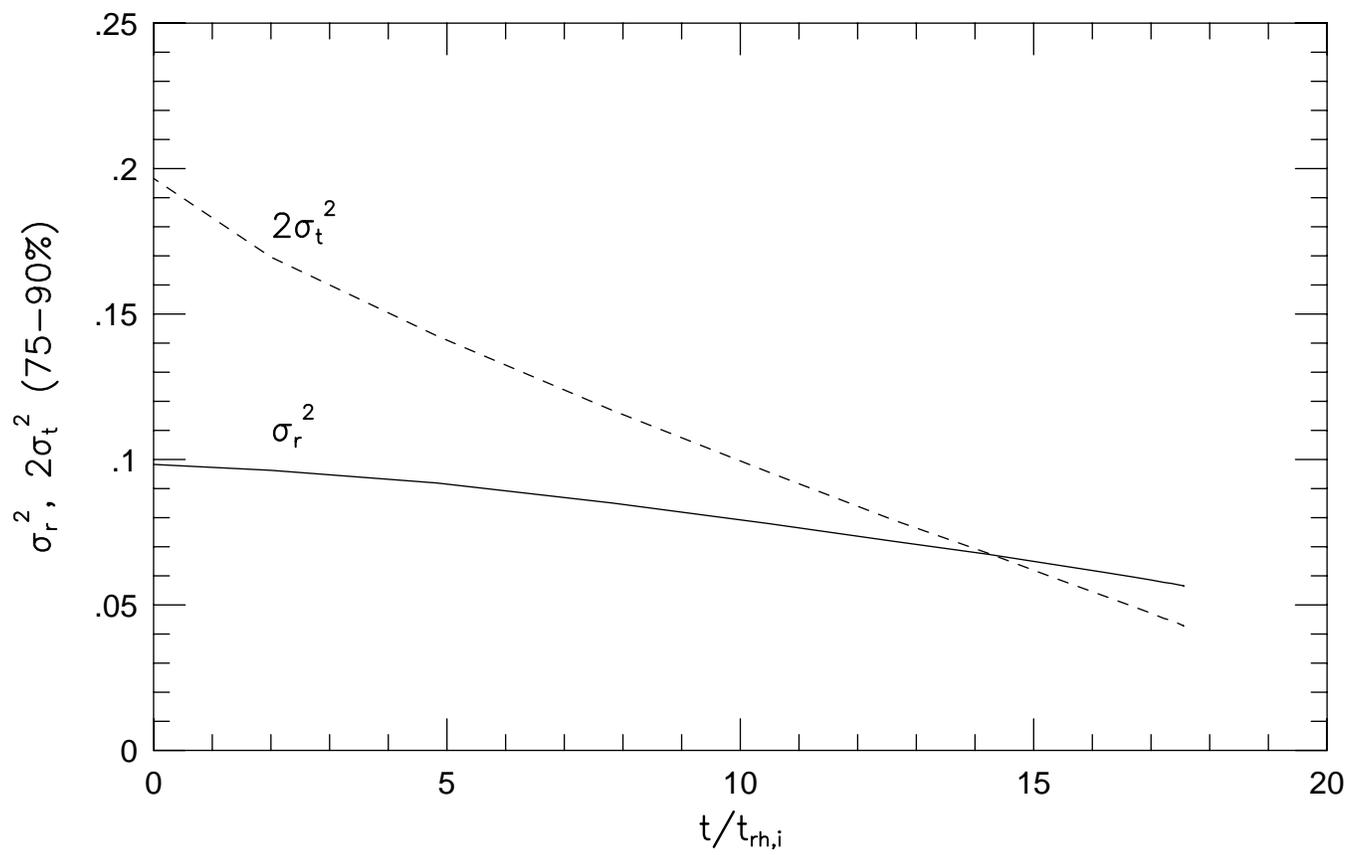
{\bf Fig. 9.}
Evolution of the velocity dispersions which are mass-averaged between 75\% and
90\% Lagrangian radii.
The solid and dashed curves are the radial and 2D tangential velocity
dispersions, respectively.

\medskip

{\bf Fig. 10.}
The distribution function $f$ at the same epoch as in figure 7 is plotted as a
function of (a) the scaled angular momentum $R$, and (b) the pericenter radius
$r_{\rm p}$, for a fixed energy $E$.
Different curves correspond to different energies $E$=0.0170, 0.0380, 0.0847,
0.188, 0.410, 0.869, 1.74, 3.14, 4.90, and 6.54, from the bottom to the top.
The central potential is $\phi(0)=7.69$.









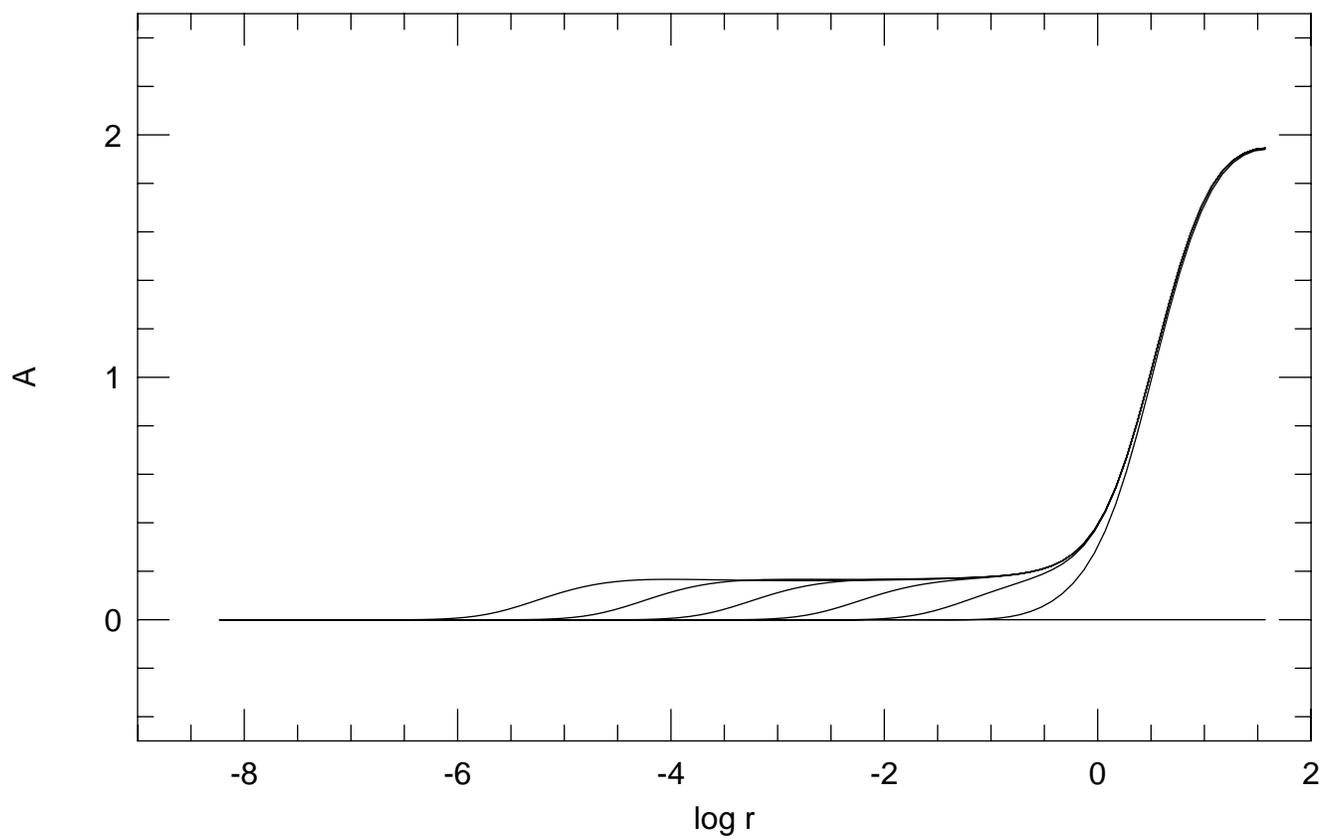
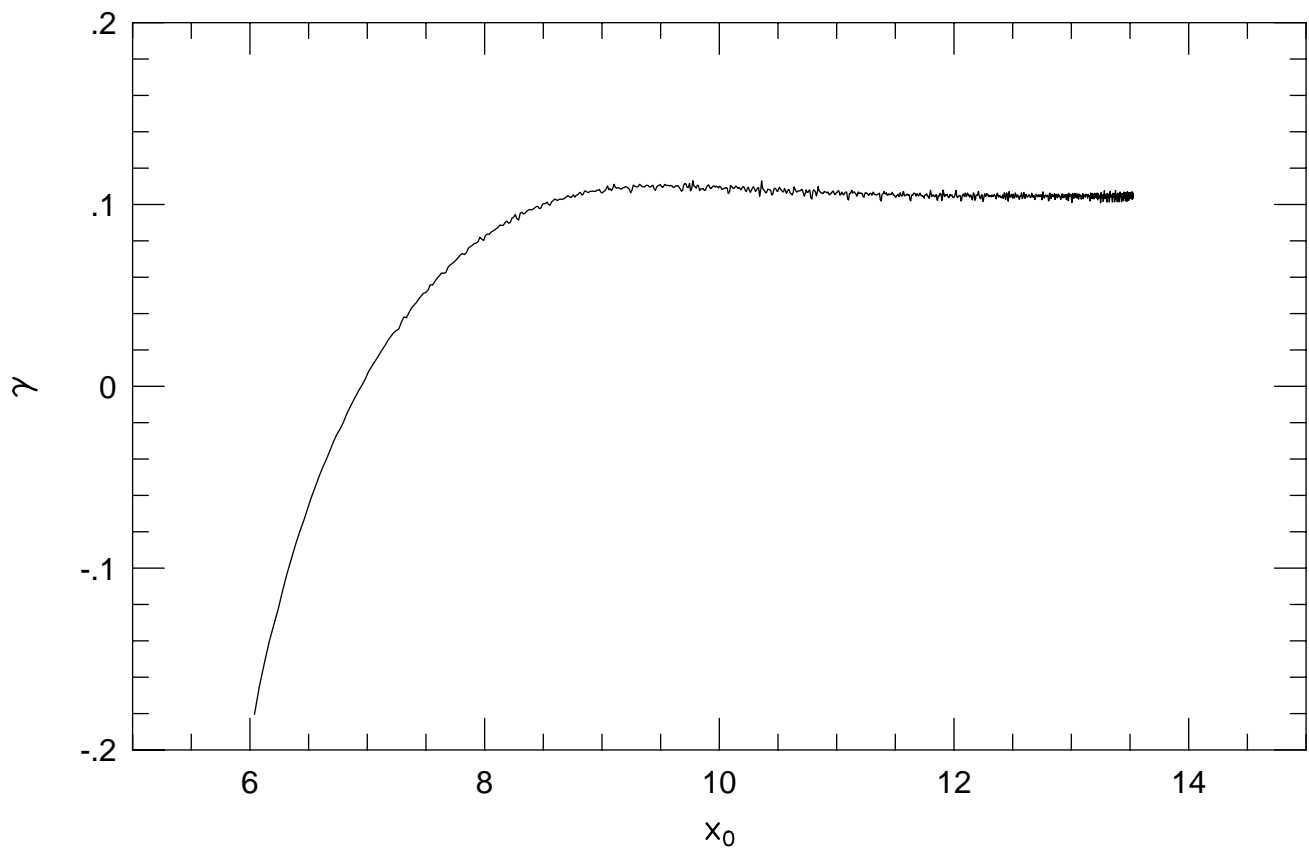
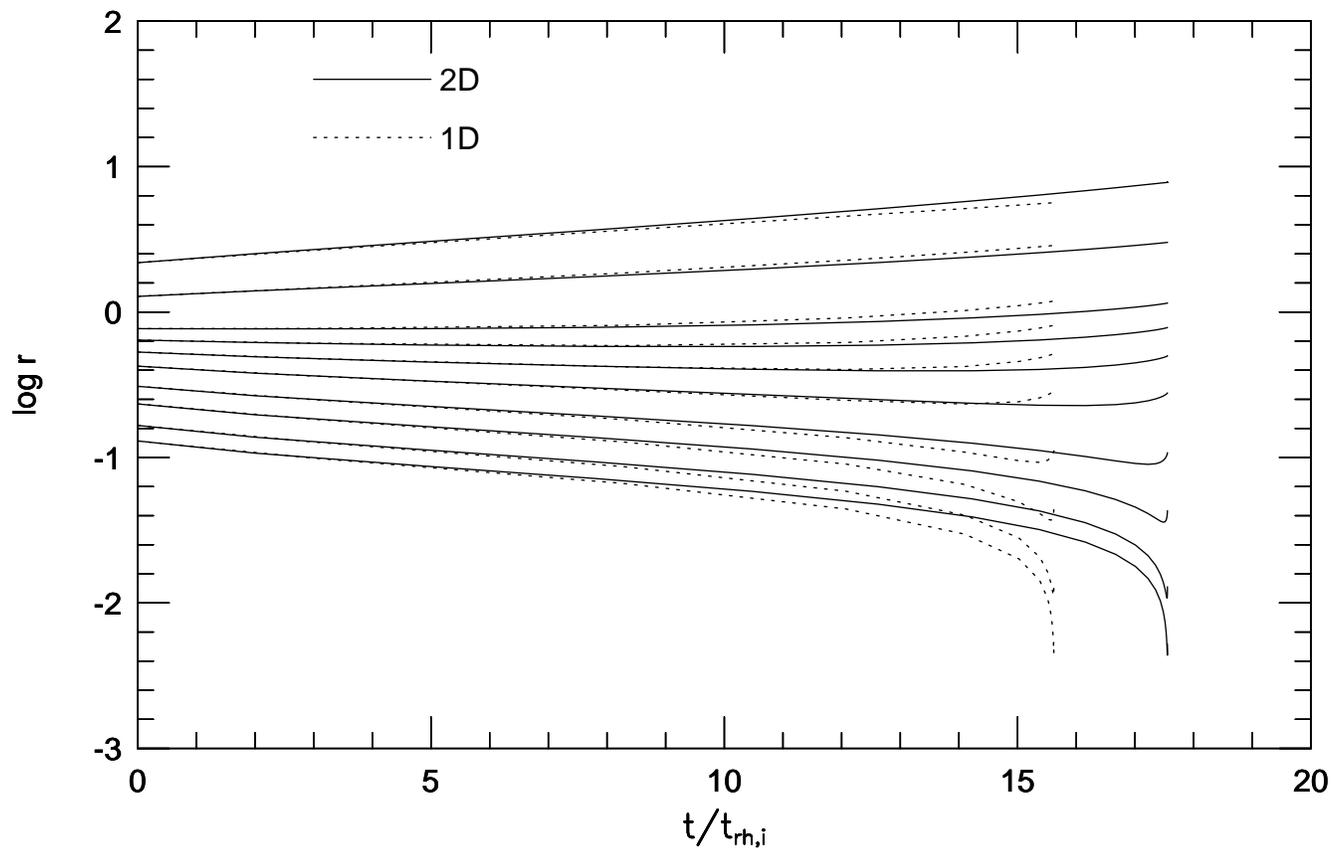
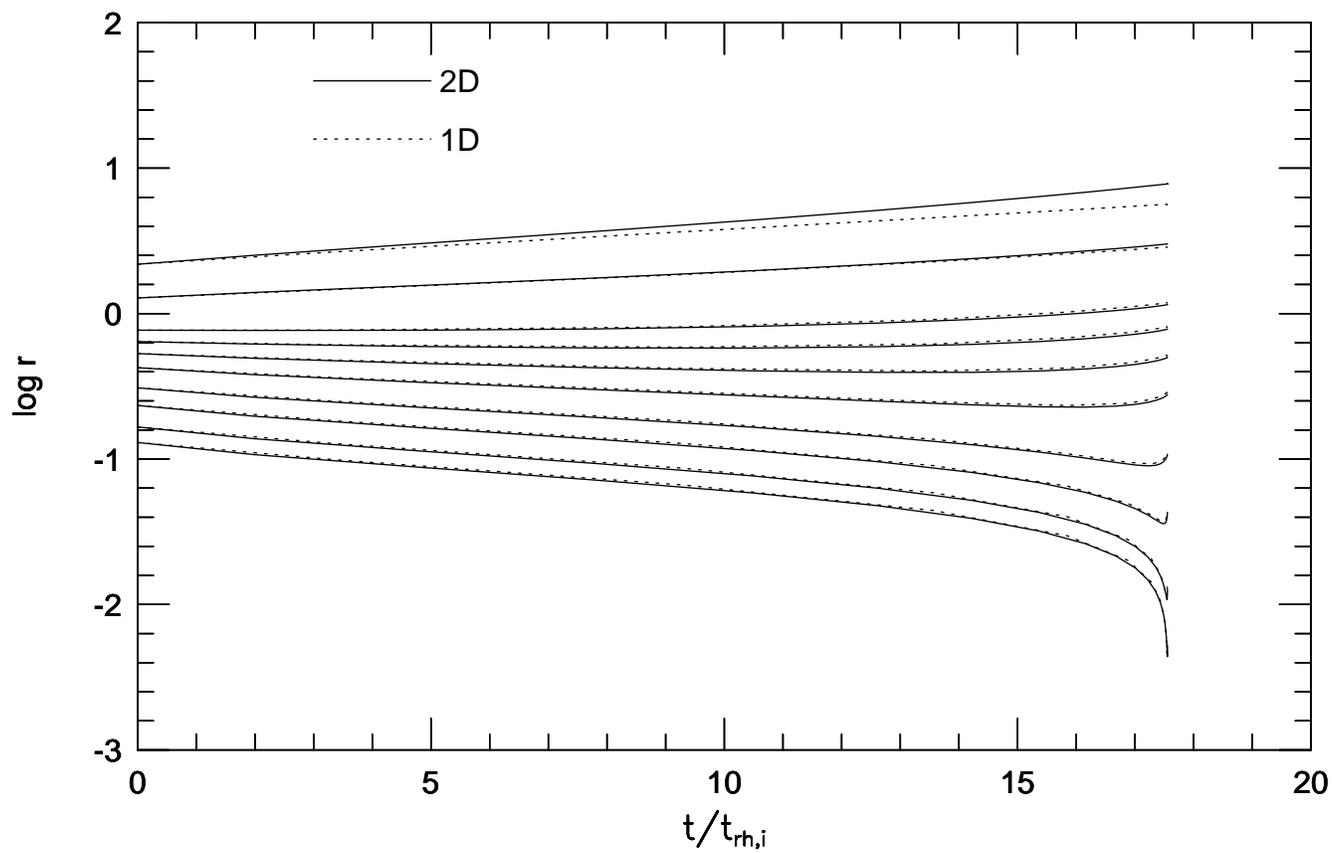
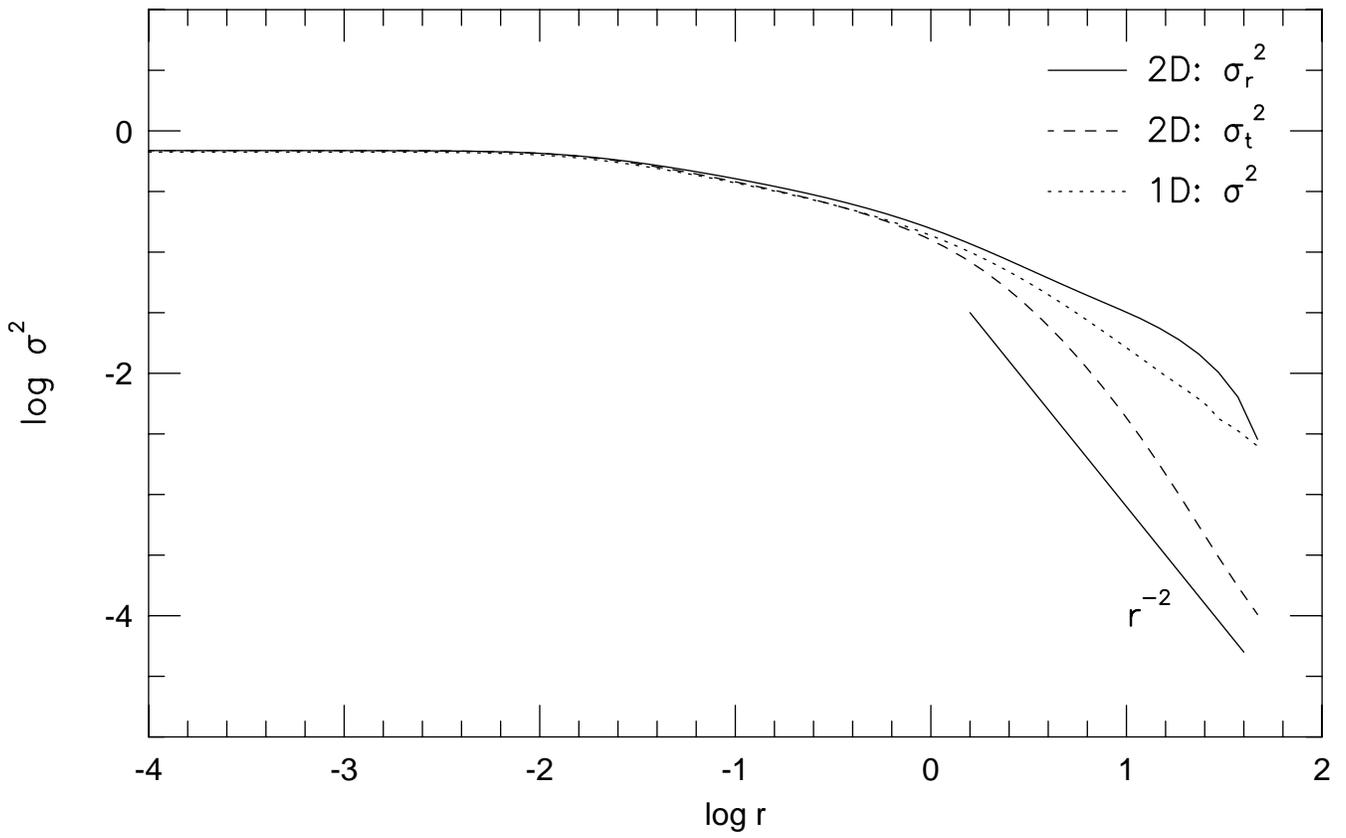
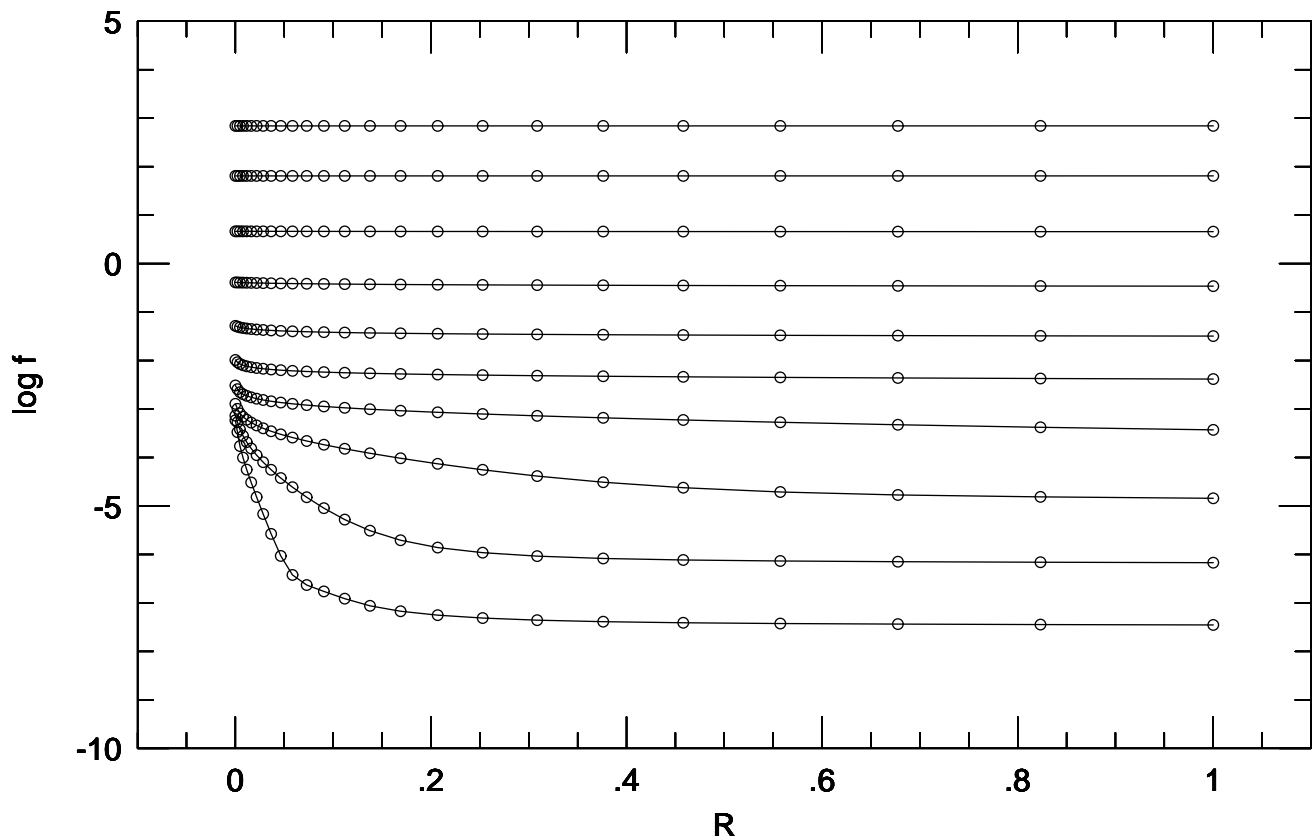
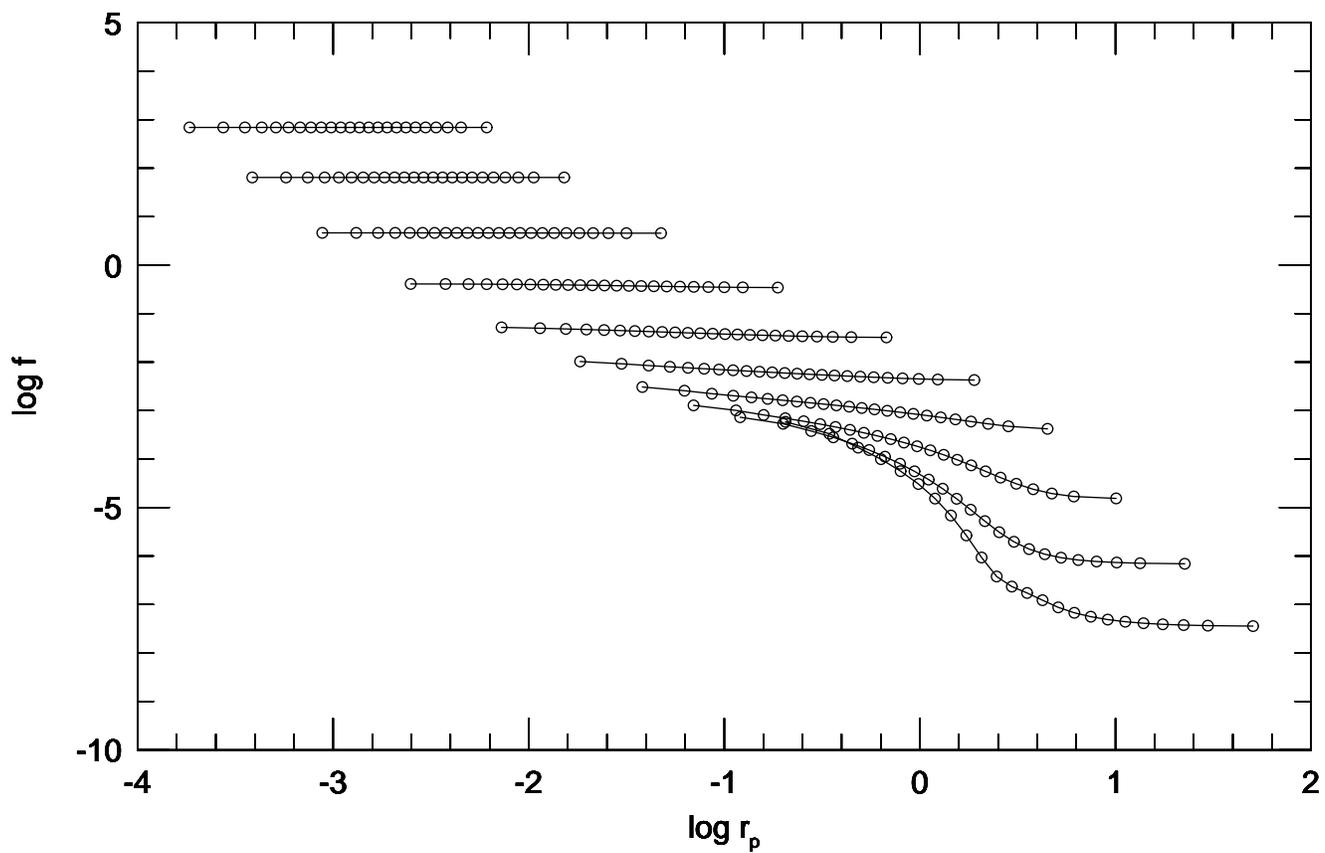
\end{document}